\documentclass[aps,prb,reprint,superscriptaddress,floatfix,longbibliography]{revtex4-2}

\usepackage{amsmath,amssymb,graphicx,bm}
\usepackage{hyperref}
\hypersetup{
  hidelinks,
  pdftitle={Ferroelectric control of the K-Gamma valley landscape in
            R-stacked bilayer WSe2},
  pdfauthor={Katsunori Wakabayashi, Souren Adhikary, Tomoaki Kameda,
            Kazuhito Tsukagoshi},
  pdfsubject={Valley physics of R-stacked bilayer WSe2},
  pdfkeywords={WSe2, R stacking, sliding ferroelectricity, K-Gamma valley,
               transconductance}}

\newcommand{\dkgTwoHPBE}{40.9}          
\newcommand{\dkgTwoHHSE}{129.2}         
\newcommand{\dkgTwoHPBEkT}{1.6}        
\newcommand{\dkgTwoHHSEkT}{5}          
\newcommand{\dWW}{6.396}                
\newcommand{\dWWR}{6.394}               
\newcommand{\alatR}{3.2806}             
\newcommand{\alatTwoH}{3.2783}          
\newcommand{\stackEdiff}{5}             
\newcommand{\mstarK}{0.386}             
\newcommand{\mstarG}{1.050}             
\newcommand{\rdosR}{3.00}               
\newcommand{\strainslopePBE}{-80.4}  
\newcommand{\strainslopeHSE}{\ensuremath{-76.6}}  
\newcommand{\strainslopefuncpct}{5}      
\newcommand{\vdwlo}{\ensuremath{-104}}  
\newcommand{\vdwhi}{27}                 
\newcommand{\vdwnone}{218}              
\newcommand{\vdwRlo}{\ensuremath{-34}}  
\newcommand{\vdwRhi}{132}               
\newcommand{\vdwdifflo}{24}             
\newcommand{\vdwdiffhi}{33}             
\newcommand{\dlo}{6.209}                
\newcommand{\dhi}{6.384}                
\newcommand{\dnone}{6.903}              
\newcommand{\starkRPBE}{17.4}           
\newcommand{\starkRHSE}{19.9}           
\newcommand{\starkRpct}{14}             

\newcommand{\tGamma}{487}               
\newcommand{\dKHSEdirect}{0.0885}       
\newcommand{\epsperpHSE}{7.23}          
\newcommand{\UFEHSE}{89.1}              
\newcommand{\UFEHSEpct}{7}              
\newcommand{\dkgshiftR}{92}             
\newcommand{\dkgdiffHSE}{31.7}          
\newcommand{\dEKtwoHHSEpair}{0.0}       
\newcommand{\dEKtwoHPBEpair}{0.015}       
\newcommand{\rWpaw}{1.455}              
\newcommand{\rSepaw}{1.164}             
\newcommand{\VPexpt}{56}                
\newcommand{\VPexptsd}{3}               
\newcommand{\VPtheo}{56}                
\newcommand{\ivdweigN}{9216}            
\newcommand{\dslab}{9.76}               
\newcommand{\fGratio}{25}              
\newcommand{\starkTwoHPBEnew}{18.1}     
\newcommand{\starkTwoHHSEnew}{20.9}     
\newcommand{\starkTwoHpct}{16}          
\newcommand{\epsperpPBER}{8.23}         
\newcommand{\epsperpHSEtwoH}{7.49}      
\newcommand{\epsperpPBEtwoH}{8.69}      
\newcommand{\DstarHSEnew}{1.01}          
\newcommand{\dratioHSE}{14}             
\newcommand{\tKtwoHbound}{0.0075}       
\newcommand{\UFE}{66.3}                 
\newcommand{\Dbuiltin}{0.102}           
\newcommand{\DvOffset}{55.9}            
\newcommand{\DvField}{0.086}            
\newcommand{\dzeroLi}{0.65}             
\newcommand{\ecrossYuan}{0.3}           
\newcommand{\strainxfer}{0.55}            
\newcommand{\epsperpWSe}{7.6}           
\newcommand{\epsperpYuan}{4.7}          

\newcommand{\polshift}{31.6}            
\newcommand{\dkgR}{72.5}                
\newcommand{\epscrossTwoH}{0.52}
\newcommand{\epscrossHSE}{1.76}         
\newcommand{\epscrossR}{0.79}           
\newcommand{\epscrossRtrans}{0.83}      
\newcommand{\Dstack}{5.6}               
\newcommand{\Vdstar}{0.19}     
\newcommand{\VdstarhBN}{6.1}            
\newcommand{\gateLi}{12}                
\newcommand{\Vdread}{20}                
\newcommand{\Dread}{0.11}               
\newcommand{\Dbuiltinext}{0.84}         

\newcommand{\Kdoubletsplit}{3.2}        
\newcommand{\GdoubletkT}{28.1}            
\newcommand{\dropfGpct}{9}             
\newcommand{\dropgdpct}{109}            
\newcommand{\KlowerA}{3.3}              
\newcommand{\KlowerB}{12.5}             

\newcommand{\dIDpct}{\ensuremath{-4.3}}              
\newcommand{\dpPct}{\ensuremath{-3.7}}               
\newcommand{\dmuPct}{\ensuremath{-0.6}}               
\newcommand{\dfGpct}{\ensuremath{+3.0}}              
\newcommand{\KlowshareZero}{14}         
\newcommand{\KlowshareFifty}{2}         
\newcommand{\dIDpctmax}{\ensuremath{-11.7}}          
\newcommand{\dmuPctmax}{-9.6}          
\newcommand{\dIDscanmax}{6}             

\newcommand{\EOT}{0.7}                  
\newcommand{\Cox}{49}                   
\newcommand{\chivpeak}{0.17}           
\newcommand{\chivpeaklegacy}{0.30}     
\newcommand{\dkgpaperII}{26}            
\newcommand{\chivpaperII}{0.19}         
\newcommand{\chivpaperIIpub}{0.20}      
\newcommand{\rN}{1.3605}               
\newcommand{\rmass}{2.7211}            
\newcommand{\mKsdh}{0.45}                
\newcommand{\gdnullspread}{20}           
\newcommand{\gdnullmeanresid}{0.24}      
\newcommand{\rNR}{1.4981}              
\newcommand{\fGpaperone}{0.21849}      
\newcommand{\pLow}{5\times10^{11}}     
\newcommand{\pHi}{1\times10^{13}}      
\newcommand{\pGhosh}{1.4}              
\newcommand{\fGRPBE}{0.095}             
\newcommand{\fGRHSE}{0.0038}            

\newcommand{\Tarrlo}{100}               
\newcommand{\Tarrhi}{500}               
\newcommand{\arrswingzero}{0.13}        
\newcommand{\arrusable}{150}            

\newcommand{\UFEdft}{83.43}                
\newcommand{\tKR}{0.0}                    
\newcommand{\tGR}{362}                    
\newcommand{\dkgRdft}{68.8}                 
\newcommand{\polshiftdft}{41.7}           
\newcommand{\dEK}{83.4}                     
\newcommand{\dEKaug}{87}                  
\newcommand{\dEKnodipole}{71}            
\newcommand{\dEKlo}{82.2}                   
\newcommand{\dEG}{727}                    
\newcommand{\socKR}{455}                  
\newcommand{\tauzG}{0.075}                
\newcommand{\tauzGneed}{0.115}            
\newcommand{\projsumK}{0.789}              
\newcommand{\projsumKhi}{0.804}            
\newcommand{\projsumG}{0.84}              
\newcommand{\vacstep}{73}                 
\newcommand{\UFEgamma}{56}                
\newcommand{\Dstardft}{1.07}             
\newcommand{\dKR}{0.0776}                
\newcommand{\epsR}{8.2}                   
\newcommand{\dKRpair}{0.0777}            
\newcommand{\dphitwoHfree}{38.9}          
\newcommand{\epsTwoHfree}{8.22}           
\newcommand{\epsRfree}{8.24}               
\newcommand{\registryagree}{0.1}           
\newcommand{\dagreepct}{5}                
\newcommand{\dekIsymhalf}{36.8}           
\newcommand{\dekIsymone}{73.6}            
\newcommand{\pfresid}{0.64}               
\newcommand{\pfresidxm}{0.58}
\newcommand{\pfresidmax}{1.6}             
\newcommand{\pfoutlier}{\ensuremath{+0.100}} 
\newcommand{\pfrange}{74}                 
\newcommand{\gdRdft}{2.0}                
\newcommand{\gcpct}{9}                   
\newcommand{\readoutLowdft}{2.4}         
\newcommand{\readoutdft}{1.2}             
\newcommand{\UFEpct}{26}                   
\newcommand{\shiftpct}{+26}               
\newcommand{\dkgdiffdft}{27.9}
\newcommand{\dkgdiffPBEpair}{27.5}      
\newcommand{\dkgdiffresid}{3.7}         
\newcommand{\polshiftgap}{8.6}           
\newcommand{\EbarKrenorm}{12.3}           
\newcommand{\GextraR}{0.04}               
\newcommand{\fGlanddown}{52}                 
\newcommand{\fGsplitup}{113}                  

\newcommand{\UFEhalf}{33.2}            
\newcommand{\UFEhalfdft}{41.7}         
\newcommand{\effpct}{100}               

\newcommand{\Gbranchshare}{10^{-13}}      
\newcommand{\tauzraw}{0.78}               
\newcommand{\tKspurious}{26}              
\newcommand{\wtopzero}{1.00}              
\newcommand{\wtopfifty}{0.89}             
\newcommand{\tKspread}{0.4}               
\newcommand{\dEKspread}{0.9}                

\newcommand{\dEKminbound}{0.035}         
\newcommand{\kTwindow}{\ensuremath{\pm0.33}}  

\newcommand{\vacpct}{14}                   
\newcommand{\Uone}{81.7}                  
\newcommand{\tone}{7.5}                   
\newcommand{\Utwo}{80.9}                  
\newcommand{\ttwo}{7.1}                   
\newcommand{\tauzone}{0.9836}             
\newcommand{\tauztwo}{0.9850}             
\newcommand{\Uproj}{81.3}                 
\newcommand{\Uprojpct}{2.5}               
\newcommand{\tauzviol}{1.1}            
\newcommand{\tauzupA}{-0.9723}            
\newcommand{\tauzloA}{0.9949}             

\newcommand{\rigidmax}{19}     
\newcommand{\rigidshift}{12.3} 
\newcommand{\Dstartrans}{0.90} 
\newcommand{\rigidzero}{3.7}  
\newcommand{\rigidboth}{6.0}  

\newcommand{\liinconsist}{16}          
\newcommand{\kTroom}{25.9}             
\newcommand{\threeway}{26}             

\newcommand{\dEKtwoHzero}{0.015}          
\newcommand{\dphitwoHzero}{0.02}           
\newcommand{\faithfulpct}{5}              
\newcommand{\dphiR}{38.8}                 
\newcommand{\dekRhalf}{38.8}                
\newcommand{\bandprec}{0.05}                 
\newcommand{\dphiRzero}{61.7}              
\newcommand{\UoverPhi}{1.35}               
\newcommand{\straincrossR}{0.86}          
\newcommand{\straincrossRown}{0.79}       
\newcommand{\strainslopeR}{-87.4}           
\newcommand{\dkgRstrainzero}{74.5}        
\newcommand{\dkgRtwo}{-94.3}              
\newcommand{\dkgRone}{-11.8}              
\newcommand{\mGstrainlo}{2.344}           
\newcommand{\mGstrainhi}{0.806}           
\newcommand{\mKstrainlo}{0.440}           
\newcommand{\mKstrainhi}{0.344}           
\newcommand{\mKR}{0.385}                  
\newcommand{\mGR}{1.154}                  
\newcommand{\mKarpes}{0.46}              
\newcommand{\mKarpeserr}{0.04}           
\newcommand{\mKcvd}{0.48}               
\newcommand{\mKarpespct}{19}             
\newcommand{\gdbandlo}{\ensuremath{-5.4}}  
\newcommand{\gdbandhi}{+2.7}             
\newcommand{\Rchannel}{7.7}              
\newcommand{\RcHalf}{0.43}               
\newcommand{\RcOne}{0.23}                
\newcommand{\RcFive}{0.03}               
\newcommand{\gdfracLo}{0.65}             
\newcommand{\gdfracHi}{0.06}             
\newcommand{\contrastFlip}{3.7}          
\newcommand{\contrastFlipRatio}{1.08}    
\newcommand{\massratioArpes}{2.5}        
\newcommand{\chiKcancelzero}{31}         
\newcommand{\chiKcancelhi}{83}           
\newcommand{\dEGzero}{726}               
\newcommand{\gdmasspct}{6}               
\newcommand{\mGstrainpct}{26}             
\newcommand{\mKstrainpct}{6.2}            
\newcommand{\massratioStrain}{4.2}        
\newcommand{\strainresid}{4.5}            
\newcommand{\strainslopeagree}{8.7}          
\newcommand{\tKlimitcanon}{0.017}        
\newcommand{\tKlimitcl}{95}              
\newcommand{\tratiobound}{2\times10^{4}} 
\newcommand{\scanpointscanon}{26}        
\newcommand{\scanresidcanon}{0.014}      
\newcommand{\scannodepoints}{6}          
\newcommand{\tKproj}{7.3}                
\newcommand{\twotKproj}{14.6}            

\newcommand{\projpoints}{five}           
\newcommand{\projdEhi}{83.1}             
\newcommand{\projdElo}{5.5}             
\newcommand{\projtauz}{0.984}            
\newcommand{\projratio}{0.091}           
\newcommand{\projratiopct}{4}            

\newcommand{\dEGtwoH}{708.0}             
\newcommand{\tGtwoHband}{354}            
\newcommand{\tGUsens}{1.0}               
\newcommand{\tGRband}{361.8}             
\newcommand{\tGbandpct}{2}             
\newcommand{\tGRbandXM}{362.2}           

\newcommand{\nbandk}{300}                
\newcommand{\dEKtwoHband}{1.5}           
\newcommand{\dEKMXband}{83.05}            
\newcommand{\dEKXMband}{83.2}            
\newcommand{\wKedge}{0.99}               
\newcommand{\wKpol}{0.998}               
\newcommand{\ftopTwoH}{0.50}             
\newcommand{\ftopR}{0.19}                
\newcommand{\wGshared}{0.46}             
\newcommand{\wGsharedb}{0.54}            
\newcommand{\bandscf}{0.4}               
\newcommand{\dEKseriesa}{44.2}           
\newcommand{\dEKseriesb}{5.5}           
\newcommand{\dEKDstar}{0.357}              
\newcommand{\dEKpast}{17.5}              
\newcommand{\tKDstar}{0.033}              
\newcommand{\tofkfar}{12.5}              
\newcommand{\tofkfarG}{7.2}              
\newcommand{\mixdk}{0.039}                
\newcommand{\wmix}{0.52/0.48}            
\newcommand{\tatmix}{6.15}                  
\newcommand{\mixhalflo}{0.15}            
\newcommand{\mixhalfhi}{0.19}            

\newcommand{\vaspold}{6.5.0}             
\newcommand{\vaspnew}{6.6.1}             

\newcommand{\dGband}{0.0590}             
\newcommand{\dKbandfit}{0.077}          
\newcommand{\dratioGK}{0.76}            
\newcommand{\dratioGKtwoH}{0.81}      
\newcommand{\dKtwoH}{0.0736}          
\newcommand{\UGband}{66.2}                
\newcommand{\UKband}{83.4}              
\newcommand{\UratiopctG}{21}            
\newcommand{\UratioGK}{0.79}            
\newcommand{\nullG}{1.1221}               
\newcommand{\nullK}{1.0745}              
\newcommand{\nullgap}{4.4}              
\newcommand{\gdturnover}{260}            
\newcommand{\dkgpredtK}{0.011}           
\newcommand{\GsepMax}{6}                
\newcommand{\screenNonDeg}{35}           
\newcommand{\screenDegen}{7}             
\newcommand{\tauzGshiftpct}{4}           
\newcommand{\gdnullpct}{1.8}            
\newcommand{\gdnulllow}{0.2}            
\newcommand{\gdnullD}{1.05}             
\newcommand{\psinglestar}{1.2}          
\newcommand{\Dsinglepct}{84}            
\newcommand{\mulayerslope}{1.84}         
\newcommand{\mulayer}{10}              
\newcommand{\mulayershift}{18.4}         
\newcommand{\uenvslope}{1.20}            
\newcommand{\uenvmev}{5}                 
\newcommand{\uenvshift}{6.0}             
\newcommand{\gdTlo}{25}                 
\newcommand{\gdThi}{500}                

\newcommand{\tslopeKM}{160}              
\newcommand{\tslopeKG}{123}              
\newcommand{\tinterKM}{-0.06}            
\newcommand{\tinterKG}{+0.13}            
\newcommand{\tinterfloor}{12}             
\newcommand{\tnoderesid}{0.11}           
\newcommand{\tfitptsKM}{8}               
\newcommand{\tfitptsKG}{4}               
\newcommand{\tfitpts}{11}                
\newcommand{\arrswingdirect}{0.02}       
\newcommand{\arrswingupper}{0.002}       
\newcommand{\dEKbandhalf}{35.0}          
\newcommand{\dEKbandone}{70.8}           
\newcommand{\bandrepro}{4}                

\newcommand{\fourtKsq}{0.28}
\newcommand{\residcanon}{0.15}
\newcommand{\residcanonR}{1.04}
\newcommand{\nfieldsR}{26}
\newcommand{\ansatzresid}{0.52}
\newcommand{\residGcanon}{0.003}
\newcommand{\tKbound}{0.0075}
\newcommand{\tKbandface}{0.18}           

\newcommand{\convmesh}{0.021}            
\newcommand{\convencut}{0.010}           
\newcommand{\convediff}{0.003}           
\newcommand{\convcell}{0.006}            
\newcommand{\convdww}{0.0009}            
\newcommand{\convmirror}{0.0019}         
\newcommand{\dkgdww}{646}                
\newcommand{\convdwwmeV}{0.6}            
\newcommand{\convmirrormeV}{1.2}         

\newcommand{\dEKnode}{0.011}             

\makeatletter
\AtBeginDocument{\immediate\write\@auxout{\string\citation{apsrev41Control}}}
\makeatother

\begin{document}

\title{Ferroelectric control of the
\texorpdfstring{$K$--$\Gamma$}{K-Gamma} valley landscape in
R-stacked bilayer \texorpdfstring{WSe$_2$}{WSe2}}

\author{Katsunori Wakabayashi}
\email{WAKABAYASHI.Katsunori@nims.go.jp}
\affiliation{Research Center for Materials Nanoarchitectonics (MANA),
National Institute for Materials Science (NIMS),
Tsukuba 305-0044, Japan}

\author{Souren Adhikary}
\affiliation{Research Center for Materials Nanoarchitectonics (MANA),
National Institute for Materials Science (NIMS),
Tsukuba 305-0044, Japan}

\author{Tomoaki Kameda}
\affiliation{Research Center for Materials Nanoarchitectonics (MANA),
National Institute for Materials Science (NIMS),
Tsukuba 305-0044, Japan}

\author{Kazuhito Tsukagoshi}
\affiliation{Research Center for Materials Nanoarchitectonics (MANA),
National Institute for Materials Science (NIMS),
Tsukuba 305-0044, Japan}
\date{\today}

\begin{abstract}
In bilayer WSe$_2$ the valence-band maxima at $K$ and $\Gamma$ lie within a few thermal energies of one another, and gating redistributes holes between a light and a heavy valley of different mobilities on a landscape that stays fixed.  We show
that rhombohedral (R) stacking moves the landscape itself.  The three-fold
symmetry of R stacking forbids interlayer tunneling at $K$ but not at $\Gamma$.  The $K$ coupling is therefore exactly zero, not merely small.  A fit that leaves the coupling free returns an upper bound of
\tKlimitcanon~meV, three orders of magnitude under $k_BT$ at room temperature, while
the $\Gamma$ coupling is \tGRband~meV.  The polarization creates a potential step between the layers and splits the
$K$ doublet.  Each $K$ state stays on a single layer, so the $K$ edge shifts
by half of that splitting, \polshiftdft~meV, which any tunneling would
reduce.  $\Gamma$ barely responds.  Two consequences follow from symmetry alone, whatever the size of that shift.  The two registries, carrying opposite out-of-plane ferroelectric polarizations, are exact mirror images along the displacement-field axis, and reversing the polarization reverses the displacement of the landscape.  Their differential-mode transconductances are equal and opposite at zero
field, where the response vanishes for an ideal nonpolar reference.  The stored polarization is read out as a reversal of sign and not as a difference of current.
Sweeping the field is a second route to the same parity.  In the ideal
symmetric limit the transconductance passes through zero at the field that
cancels the polarization, \Dstardft~V\,nm$^{-1}$, at a single temperature.  Both routes are limited by the symmetry of the device more than by the band
structure.  Stacking also renormalizes the landscape, so the polar one is not
a translated copy of the nonpolar reference.  Absolute
offsets follow the exchange--correlation functional.  The symmetry relations
do not.
\end{abstract}

\maketitle

\section{Introduction}
\label{sec:intro}

The valley degree of freedom in two-dimensional (2D) transition-metal
dichalcogenides (TMDCs) is being developed as a basis for
devices~\cite{Xiao2012_valley,Schaibley2016_valleytronics,Manzeli2017_TMDreview,Geim2013_vdW,Lai2023_valleytronic}.
A monolayer has two valleys, $K$ and $K'$, which are time-reversed partners
and are degenerate in zero magnetic field.  Since the monolayer has no
inversion center, the two carry opposite Berry curvature and absorb
opposite circular polarizations of
light~\cite{Xiao2012_valley,Xu2014_spinpseudo}.  Spin--orbit coupling splits
the valence band at $K$ by several hundred meV and locks the spin to the
valley~\cite{Zhu2011_SOC}.  Adding layers brings a third valley, $\Gamma$ at
the zone center, into competition for the top of the valence band.  In
WSe$_2$ that maximum moves from $K$ towards $\Gamma$ as layers are added,
because the layers
hybridize~\cite{Zhao2013_WSe2layers,Yeh2015_WSe2muARPES,Wilson2017_muARPES},
as they do in the group-VI dichalcogenides generally~\cite{Komsa2012_indirect,Zhang2014_MoSe2}.
The bilayer sits where the two are nearly degenerate.

\begin{figure*}[tb]
  \includegraphics[width=\textwidth]{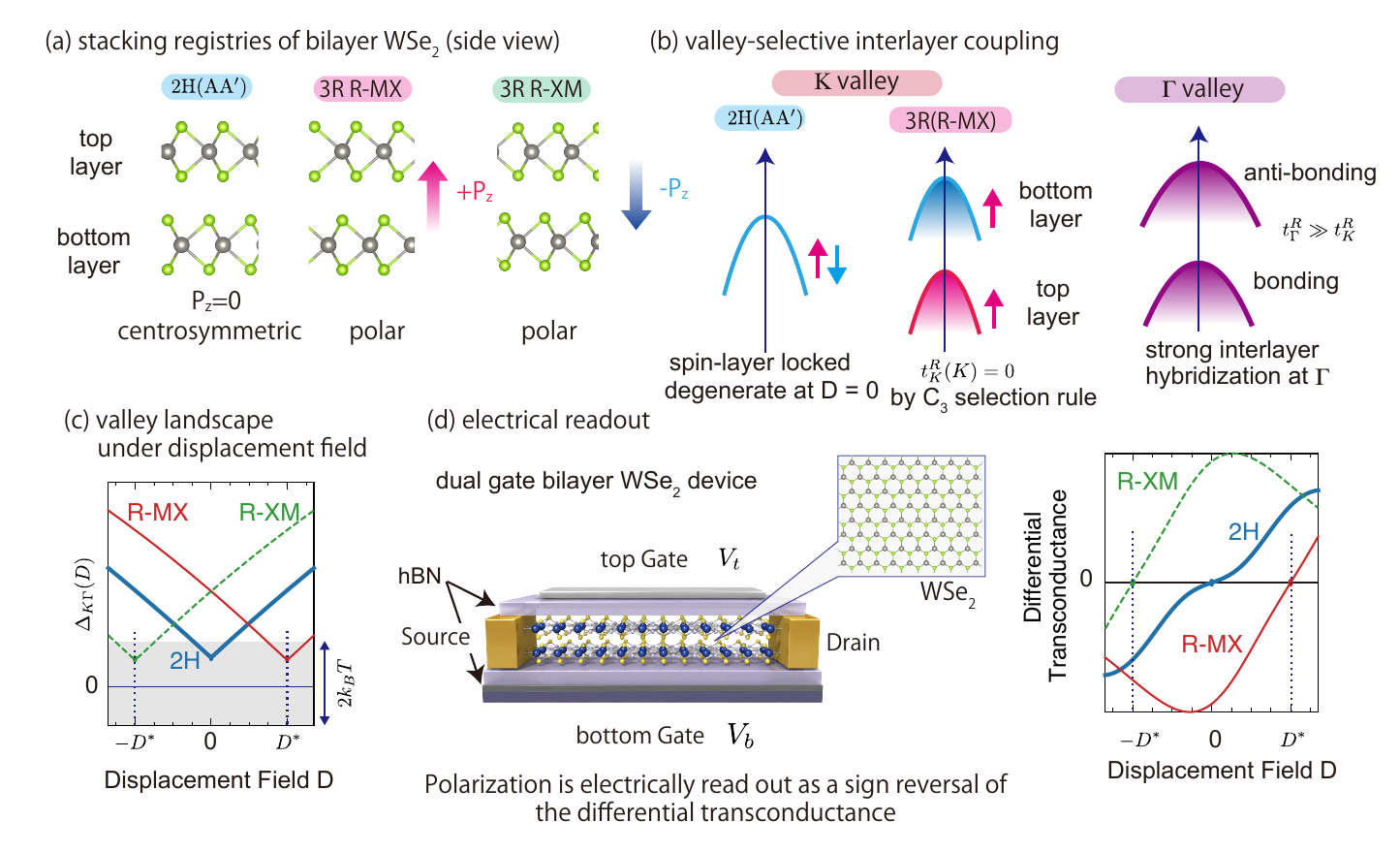}
  \caption{%
    \textbf{The valley landscape depends on the stacking and moves with a
    displacement field.}
    (a)~The three registries of bilayer WSe$_2$ in side view, defined in
    Sec.~\ref{sec:registry}.  2H (AA$'$) is centrosymmetric and has $P_z=0$;
    the two R registries are translations of one another and carry opposite
    polarizations, $P_z^{\rm MX} = -P_z^{\rm XM}$.  The absolute direction
    drawn is a first-principles result and not a convention, from two independent determinations that agree (Appendix~\ref{app:rparams}), and $s=-1$ for R-MX in Eq.~(\ref{eq:Delta}) with $P_z>0$.
    (b)~The mechanism of Sec.~\ref{sec:calibration}.  In 2H the $K$ pair is
    degenerate at $D=0$ and carries opposite spins on opposite layers.  In R stacking both branches have the same spin.  Spin protects nothing, and the three-fold selection rule forbids the hopping instead,
    $t_K^{\rm R}(K)=0$, leaving pure layer states.  The rule holds for both
    polar registries; the panel is drawn for R-MX, whose $K$ edge lies on
    the bottom layer, and R-XM is the layer-exchanged mirror of R-MX.
    At $\Gamma$ the layers
    hybridize, $t_\Gamma \gg t_K$, into bonding and antibonding combinations shared between them, and a field acting on the layer degree of freedom shifts the two valleys by different amounts.
    (c)~The valley offset, drawn schematically.  The two polar
    registries are exact mirrors of one another in the field axis,
    Eq.~(\ref{eq:mirror}), with minima displaced by the polarization to
    $\pm D^{*}$, the field that cancels the polarization; that displacement is the
    paper's central claim.  The
    nonpolar curve is not a third member of the same family.  The shape
    matches to \rigidboth~meV, but only after both a horizontal translation
    and a vertical offset are allowed, because changing the stacking both
    displaces and renormalizes the landscape
    (Appendix~\ref{app:transfer}).  The shaded band is the thermal window
    $|\Delta_{K\Gamma}| < k_BT$ at 300~K, of width $2k_BT$ as the arrow
    marks, in which both valleys are occupied.
    (d)~The same axis, read electrically.  In the dual-gated device the two
    gate voltages enter only through their sum and difference, $V_c$
    controlling carrier accumulation and $V_d$ the field,
    $D=\varepsilon_{ox}(V_t-V_b)/2t_{ox}$ for symmetric gates
    [Eq.~(\ref{eq:Dvd})].  In the symmetric model of
    Fig.~\ref{fig:gdnull}(a), which gives both valleys one dipole length,
    the differential transconductance vanishes where each landscape is centered [Eq.~(\ref{eq:gdnull})], and the stored polarization is read out as the sign of $g_d$.
    Panels~(c) and~(d) are schematic.  The field axis has no volts because the conversion brings in the gate dielectric --- the same
    $D^{*}$ is $V_d = \Vdstar$~V on the scaled stack and $\VdstarhBN$~V on
    20~nm hBN --- and the shapes are drawn to show the relations, not to be
    read off.  Figures~\ref{fig:stark} and~\ref{fig:gdnull} give the
    calibrated curves.%
  }
  \label{fig:registries}
\end{figure*}

The two valleys differ strongly in effective mass.  The nonpolar bilayer
gives $m^*_K = \mstarK\,m_0$ and $m^*_\Gamma = \mstarG\,m_0$, with $m_0$ the
free-electron mass;
Shubnikov--de Haas oscillations in
gated monolayer and bilayer WSe$_2$ give the lighter of the
two~\cite{Fallahazad2016_mass}.  Moving holes between the valleys therefore
changes the effective mobility without changing the total carrier
density~\cite{Wickramaratne2014_indirect,Jin2014,Sohier2019_valleyEng}.
Experiment shows that such a redistribution is real and can be driven by a
gate: in hole-doped trilayer WSe$_2$ the $\Gamma$ and $K$ populations have
been tuned through one another by density and displacement
field~\cite{Movva2018_valleyPRL}.  A perpendicular field shifts the valleys themselves as well.  First-principles
calculations of a six-layer WSe$_2$ slab place the valence-band maximum at
$\Gamma$ below \ecrossYuan~V\,nm$^{-1}$ and at $K$ above
that field~\cite{Yuan2013_zeeman}.  In that calculation the field is applied from
outside.  A stacking that breaks inversion symmetry produces an interlayer
potential already present at zero applied field.  We show
that this potential displaces the \emph{valley landscape} --- the
$K$--$\Gamma$ offset $\Delta_{K\Gamma}$ as a function of applied field ---
along the field axis, in a direction that depends on how the layers are
stacked.
Figure~\ref{fig:registries} lays out the argument: the registries
in panel~(a), the valley-selective interlayer coupling that makes the
displacement possible in~(b), the displaced landscape in~(c), and the
electrical readout that follows in~(d).

Earlier studies treated $\Delta_{K\Gamma}$ as a fixed design variable in
gate-all-around transistors~\cite{PaperI_valleyGAA} and identified the
occupation signature in the
transconductance~\cite{PaperII_transconductance}.  Here the chemical
potential and the landscape itself both move.

Rhombohedral (R-stacked) bilayers change that premise.  When two monolayers are stacked with
parallel orientation, inversion symmetry is broken and interlayer charge
transfer produces an out-of-plane polarization that can be reversed by
interlayer sliding.  The mechanism was predicted for binary compound
bilayers generally~\cite{LiWu2017_sliding} and has since been realized in
bilayer WTe$_2$~\cite{Fei2018_WTe2}, in parallel-stacked hexagonal boron
nitride~\cite{VíznerStern2021_sliding}, in marginally twisted transition
metal dichalcogenides~\cite{Weston2022_marginal}, and in parallel-stacked
bilayer dichalcogenides including WSe$_2$ itself, whose built-in interlayer
potential has been measured~\cite{Wang2022_interfacialFE}.
In the rhombohedral bilayer the polarization is switched by the motion of
domain walls~\cite{Yang2024_domainwall}.  For
WSe$_2$ specifically, optical spectroscopy of R-stacked bilayers places
the interfacial potential at \UFE~meV, with a
built-in field of \Dbuiltin~V\,nm$^{-1}$~\cite{Li2026_RWSe2}.  That energy is
comparable to the $K$--$\Gamma$ offset itself, which our own calculations
place between \dkgTwoHPBE{} and \dkgTwoHHSE~meV in the nonpolar bilayer,
according to the exchange--correlation functional.

The $K$-valence state of a bilayer is built from in-plane W $d$ orbitals and
couples only weakly between the layers; in the polar registries, as
Sec.~\ref{sec:calibration} shows, symmetry removes that coupling outright.  The $\Gamma$-valence state has
strong out-of-plane character and forms bonding and antibonding
combinations split by \dEGtwoH~meV in the nonpolar bilayer.  Thus, a layer potential
polarizes the $K$ states almost completely and leaves the $\Gamma$ states
nearly untouched.  The two valence edges shift by different amounts, and the difference
changes $\Delta_{K\Gamma}$.

That orbital asymmetry is
established~\cite{Komsa2012_indirect,Zhao2013_WSe2layers,Zhang2014_MoSe2,Yeh2015_WSe2muARPES,Wilson2017_muARPES}.  We connect that contrast to transport in three steps: calibrating the
Stark response on the \emph{nonpolar} 2H bilayer, where no polarization
exists to produce the contrast (Sec.~\ref{sec:calibration}); applying the
response to the polar R-stacked landscape (Sec.~\ref{sec:A1}); and
propagating the resulting displacement through layer-resolved carrier
statistics to the differential transconductance
(Secs.~\ref{sec:thermo} and~\ref{sec:transport}).  The technical core of
all three is the symmetry-enforced node of the $K$-valley interlayer
hopping of the polar stacking.
Symmetry forces a node in that hopping exactly at $K$, and establishing the node is the
technical core of the paper.

The status of the $K$/$\Gamma$ ordering in R-stacked WSe$_2$ is itself
unsettled.  The optical study above assigns the valence-band edge to
$K$~\cite{Li2026_RWSe2}, whereas angle-resolved photoemission on
quasi-free-standing rhombohedral bilayers grown on W(110) reports a $\Gamma$ valence-band maximum~\cite{Chapuis2026_3RW110}. Photoemission on chemical-vapour-deposited 3R bilayers resolves hole pockets at both points, and the calculations in that work place the maximum at $\Gamma$ with a hybrid functional and at $K$ with a quasiparticle correction~\cite{Mahmoudi2023_3Rbilayer}.  We do not settle that question here, and show that both assignments lie within the range plausible
strain and interlayer separation can produce, which our own calculations
span as well.

The calculated offset changes with the exchange--correlation functional and
with the dispersion treatment.  Our own converged calculations place the
$K$ edge of the nonpolar bilayer above the $\Gamma$ edge by
$\Delta_{K\Gamma} = \dkgTwoHPBE$~meV at the PBE-D3(BJ) level and by \dkgTwoHHSE~meV at
HSE06~\cite{Heyd2003_hse,Heyd2006_hse_erratum}, about $\dkgTwoHPBEkT\,k_BT$ in one case
and about $\dkgTwoHHSEkT\,k_BT$ in the other at room temperature
($k_BT = \kTroom$~meV~\cite{Tiesinga2021_codata}).  Thus, the $\Gamma$ occupation is uncertain by a large factor.  We compared four
dispersion treatments~\cite{PaperI_valleyGAA}, all within density-functional
theory (DFT): DFT-D2, optB88-vdW, and DFT-D3 with
Becke--Johnson and with zero damping.  Across them $\Delta_{K\Gamma}$ of the
2H bilayer spans \vdwlo~meV to $+$\vdwhi~meV, changing sign on the way,
while the interlayer W--W distance spans \dlo~\AA{} to \dhi~\AA.
Omitting dispersion altogether returns an offset of $+$\vdwnone~meV, but at
$\dnone$~\AA{}, half an \AA ngstrom beyond any bound bilayer.
Section~\ref{sec:calibration} gives the settings, which differ only in that
correction.  The offset behaves as a structure-dependent
landscape variable and not as a material constant, consistent with the
pressure dependence reported for \emph{moir\'e} bilayer WSe$_2$~\cite{Brzezinska2025_pressure}.  We organize the results below around displacements of the valley landscape.

Throughout we separate three classes of statement: results from direct
calculation, results protected by symmetry, and predictions that depend on
the model, on the functional or on the device geometry.  The Conclusion
sorts the results under the first two headings and states the experiment
they specify.  The third class is the subject of Sec.~\ref{sec:discussion}
and Appendix~\ref{app:robust}, and those predictions are quoted as bands
wherever they enter.

One boundary should be drawn at the outset.  ``Control'' here means the control
of the valley landscape by a polarization state that is taken as given, and
the electrical read-out of that state.  We do not compute the switching path,
the barrier separating the two registries, the coercive field or the time
required to write one state into the other.  Those belong to domain-wall
kinetics, which Sec.~\ref{sec:discussion} returns to.

A jointly submitted companion Letter states the symmetry principle and the
parity results that follow~\cite{CompanionLetter}; this
self-contained Article gives the derivation, the calibration against first
principles, and the limits.

\section{Layer-resolved two-valley model}

The model below keeps two states per valley, one on each layer, and
nothing else.  Every result below traces back to a parameter or to a
symmetry.  Dispersion enters only through the effective masses.

\subsection{Registry notation}
\label{sec:registry}

We define the notation for the two polar registries explicitly.  Figure~\ref{fig:registries}(a) shows the three
structures side by side.  The
difference between 2H and R stacking is one of orientation.  A 2H (AA$'$)
bilayer is made by rotating the second layer through $180^\circ$.  Every R registry, of the 3R type, is made by
sliding the second layer in the plane, with both layers keeping their
orientation~\cite{GongZ2013_magnetoelectric}.

Write $M$ for W and $X$ for Se.  Each layer is a Se--W--Se sandwich whose
two Se share an in-plane coordinate.  Thus, the slide leaves each layer
intact and only decides which atom of the upper layer sits above which
atom of the lower one.  The two names record that alignment.  In \textbf{R-MX} a
top-layer W sits directly above a bottom-layer Se; in \textbf{R-XM} a
top-layer Se sits above a bottom-layer W.  Exchanging the layers maps one onto the other, and the two carry opposite
out-of-plane polarization; both are $P3m1$, point group $C_{3v}$.  In 2H
both alignments occur at once, the bilayer is centrosymmetric, $P\bar{3}m1$
with point group $D_{3d}$, and the polarization vanishes.

\subsection{Hamiltonian}

For each valley $\nu \in \{K,\Gamma\}$ we work in the layer
pseudospin space spanned by the top and bottom monolayer states, in that
order, with Pauli matrices $\tau$ and identity $I$, so that $\tau_z = +1$ on
the top layer.  The valence Hamiltonian is
\begin{equation}
H_\nu^{(s)}(D) = \bar E_\nu \, I + t_\nu \tau_x
  + \tfrac{1}{2}\Delta_\nu^{(s)}(D)\, \tau_z ,
\label{eq:H}
\end{equation}
with the layer-potential asymmetry $\Delta_\nu^{(s)}(D)$, the difference
between the valley-$\nu$ on-site energies of the two layers,
\begin{equation}
\Delta_\nu^{(s)}(D) = E_\nu^{\rm top} - E_\nu^{\rm bot}
  = s\,U^\nu + e\,d_\nu D + U_{\rm env} .
\label{eq:Delta}
\end{equation}
Here $\bar E_\nu = \tfrac{1}{2}(E_\nu^{\rm top} + E_\nu^{\rm bot})$ is the mean
of that same pair, and $t_\nu \geq 0$ is the interlayer hopping.  The
sum $\bar E_\nu + t_\nu$ is the band edge of the unpolarized bilayer at zero
field.  The diagonal entries of Eq.~(\ref{eq:H}) are $\bar E_\nu \pm \Delta_\nu^{(s)}/2$,
and $\Delta_\nu^{(s)}$ is the whole interlayer step.  The valley index is there because the two valleys see
different steps, and both differ from the electrostatic potential
difference between the two tungsten planes.

In Eq.~(\ref{eq:Delta}), $U^\nu > 0$ is the interlayer potential step that
the polarization produces at valley $\nu$, and $s$ is the registry index,
which records the polarization state through the sign with which that
step enters.  The value is $-1$ for R-MX, $+1$ for R-XM and $0$ for the
nonpolar 2H reference.  In
2H stacking the polarization vanishes by inversion symmetry,
and $U^\nu = 0$.
Exchanging the two layers reverses $s$.  The layers
screen the applied field.  The states feel $e d_\nu D$ between
them, with $d_\nu$ much shorter than the layer spacing.

The third term, $U_{\rm env}$, is a registry-independent layer asymmetry of
environmental origin, already present at $D = 0$ because the two faces of a
bilayer in a device see different surroundings.  Two properties keep $U_{\rm env}$ out of the $e d_\nu D$ term.  Balanced gates leave the offset in place, since the
stack produces the offset and the gates do not, and so does a reversal of
the polarization: the registry index multiplies $U^\nu$ alone.  That second
asymmetry makes a deliberate $U_{\rm env}$ a design option as well as a
defect, and Sec.~\ref{sec:symmetry} returns to that option.  Every
calculation reported here is for a freestanding slab, and
$U_{\rm env} = 0$ throughout.  Appendix~\ref{app:robust} gives the sources
of such an offset in a device, the valley index a complete treatment would
carry, and the error on the read-out field that follows.

The registry enters Eq.~(\ref{eq:Delta}) in two distinct ways: through the
polar term $s\,U^\nu$, which is odd under layer exchange and vanishes
in 2H, and through the layer-symmetric parameters $\bar E_\nu$ and $t_\nu$,
which are even and differ between stackings.

Energies are in eV and potentials in V
throughout.  The elementary charge is then unity in the numerical work, and
we drop the charge from expressions such as $\partial E/\partial D$ below.
Equation~(\ref{eq:Delta}) keeps the charge for dimensional clarity.  The field variable $D$ is the
out-of-plane displacement field expressed in V\,nm$^{-1}$, that is, $D$ here
denotes the quantity written $D/\varepsilon_0$ in SI units, and every axis
labeled $D$ in the figures below is that quantity.  The dual-gated
measurement of $\Gamma$ and $K$ populations in trilayer WSe$_2$ uses that
quantity and those units, written
$E = |C_tV_t - C_bV_b|/2\varepsilon_0$~\cite{Movva2018_valleyPRL}.  The
first-principles field scan of Sec.~\ref{sec:calibration} applies $D$
itself, as an external perpendicular field on a slab, and the model
calibrates directly against that scan.

The calculations reported here determine which registry gets which sign.  They place the $K$ valence edge of R-MX on the \emph{bottom} layer, and the
vacuum-level step of that calculation has the matching sign.  The two
indications are independent.  The assignment enters below only as the
labelling of which registry sits at which field.

Diagonalizing Eq.~(\ref{eq:H}) gives the two layer-derived subbands
\begin{equation}
E_{\nu,\pm} = \bar E_\nu \pm
  \sqrt{t_\nu^2 + \left(\Delta_\nu/2\right)^2},
\label{eq:eigen}
\end{equation}
of which the upper is the valence-band edge.  Here and below $\Delta_\nu$
abbreviates $\Delta_\nu^{(s)}(D)$ of Eq.~(\ref{eq:Delta}), with the registry
and the field left implicit.  The single quantity that governs
the field response is the layer polarization
\begin{equation}
\zeta_\nu \equiv \frac{\partial E_{\nu,+}}{\partial \Delta_\nu}
  = \frac{1}{2}\,\frac{\Delta_\nu/2}{\sqrt{t_\nu^2+(\Delta_\nu/2)^2}}
  = \frac{1}{2}\langle \tau_z\rangle ,
\label{eq:zeta}
\end{equation}
where the second equality is the Hellmann--Feynman theorem applied to
$\partial H_\nu/\partial\Delta_\nu = \tau_z/2$.  The top-layer weight of the
band edge is $w_{\rm top} = 1/2 + \zeta_\nu$, and the field shifts that edge
at $\partial E_{\nu,+}/\partial D = d_\nu \zeta_\nu$, the Stark slope.

Equation~(\ref{eq:zeta}) makes the mechanism explicit.  For a layer-decoupled
valley, $|\Delta_\nu| \gg t_\nu$, the polarization saturates at
$|\zeta_\nu| = 1/2$, with the sign of $\Delta_\nu$ naming the layer the edge occupies: the state sits entirely on that layer, and the edge then
shifts linearly in the layer potential, at the steepest rate the two-level
form allows.  For a
strongly hybridized valley, $|\Delta_\nu| \ll t_\nu$, the state is shared
equally between layers, $\zeta_\nu \to 0$, and the leading response is
quadratic, $E_{\nu,+} - \bar E_\nu - t_\nu \simeq (\Delta_\nu/2)^2/2t_\nu$.  Thus, a contrast in
$t_\nu$ between the two valleys converts a common layer potential
into a differential valley shift.  We call this a valley-selective Stark
response, and Sec.~\ref{sec:calibration} measures how large the response is.

\subsection{Symmetry constraints}
\label{sec:symmetry}

Two exact statements follow from Eqs.~(\ref{eq:H})--(\ref{eq:Delta}) and hold
independently of any parameter value.

First, in an ideal top--bottom symmetric environment ($U_{\rm env}=0$) the two
registries are related by layer exchange, and any observable $X$ that does
not distinguish the layers obeys
\begin{equation}
X_{\rm MX}(D) = X_{\rm XM}(-D).
\label{eq:mirror}
\end{equation}
Any violation of Eq.~(\ref{eq:mirror}) in a calculation is an error in the
setup, and Appendix~\ref{app:rparams} reports the residuals as one of the four
checks there.

Second, Eq.~(\ref{eq:mirror}) gives
$X_{\rm MX}(0) = X_{\rm XM}(0)$ at $D=0$, and reverses the sign on
differentiation.  The two registries share every such observable at
zero field and have opposite slopes of that observable.  The read-out reads the
slope.  Section~\ref{sec:transport} works this out for the
differential-mode transconductance.  At $D = 0$ the two polar registries
have opposite slopes, and the nonpolar reference has a zero that symmetry enforces.

Equation~(\ref{eq:mirror}) assumes $U_{\rm env} = 0$.  Layer exchange
reverses $U_{\rm env}$ along with the registry, while rewriting a fabricated device leaves the gate stack alone, and $U_{\rm env}$ stays.  A device built with $U_{\rm env} \neq 0$ lies outside
Eq.~(\ref{eq:mirror}), and the two registries differ in the current itself at
$D = 0$.

\section{Valley-selective interlayer coupling in the nonpolar bilayer}
\label{sec:calibration}

At $\Gamma$ the two layers of bilayer WSe$_2$ hybridize across several
hundred meV.  At $K$ the mixing is strongly suppressed by spin--layer locking in 2H and forbidden outright by the three-fold rotation in R.  We establish that
contrast on the \emph{nonpolar} 2H bilayer, where inversion symmetry requires $U^\nu = 0$, i.e., where no polarization exists to produce the contrast.
Section~\ref{sec:A1} measures both couplings in R stacking, where
the polarization is present.

\subsection{The 2H bilayer as a nonpolar reference}

Owing to the inversion symmetry in 2H, $U^\nu$ in Eq.~(\ref{eq:Delta})
vanishes.  In addition, $U_{\rm env}$ is absent, because the calculations
here are for a freestanding slab.  The only layer asymmetry left is the one the applied
field creates,
$\Delta_\nu = e\,d_\nu D$, and that asymmetry raises both valence edges.  The
calculations below supply their separation,
$\Delta_{K\Gamma} = E_{K,+} - E_{\Gamma,+}$ in the notation of
Eq.~(\ref{eq:eigen}).  The layer-averaged energies $\bar E_\nu$ are the same
at every point of the scan.  Subtracting the zero-field separation
removes them and leaves only the field response:
\begin{equation}
\begin{aligned}
\Delta_{K\Gamma}(D) = \Delta_{K\Gamma}(0)
 &+ \left[ \sqrt{t_K^2 + (e d_K D/2)^2} - t_K \right] \\
 &- \left[ \sqrt{t_\Gamma^2 + (e d_\Gamma D/2)^2} - t_\Gamma \right] .
\end{aligned}
\label{eq:dkgD}
\end{equation}
In 2H the two valleys enter with opposite signs.  The field shifts $\Delta_{K\Gamma}$ by the difference between their responses
alone: if
$t_K = t_\Gamma$ and $d_K = d_\Gamma$, the two brackets are identical and
$\Delta_{K\Gamma}$ is independent of $D$.  Thus, a scan of
the nonpolar bilayer measures the contrast between the valleys,
and not either valley by itself.

\begin{figure}[tb]
  \includegraphics[width=\columnwidth]{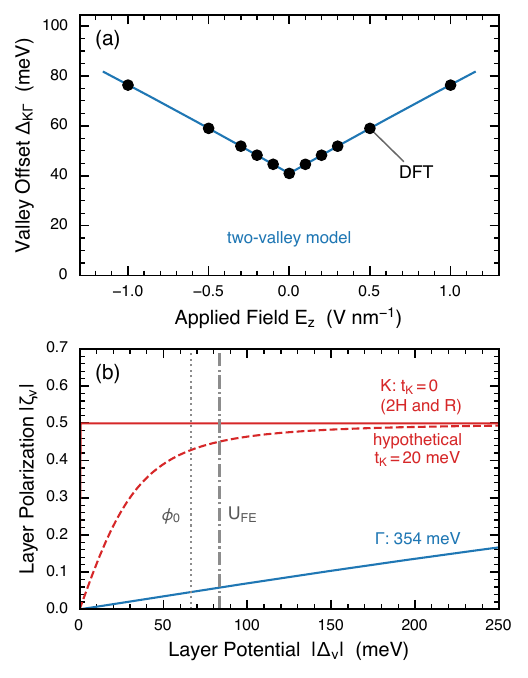}
  \caption{%
    Calibration of the layer-resolved two-valley model against
    first-principles data for the nonpolar 2H bilayer.
    (a)~$\Delta_{K\Gamma}$ as a function of applied out-of-plane field: symbols are the
    calculated values, the curve is Eq.~(\ref{eq:dkgD}) with every parameter
    read off a splitting of the same scan and nothing fitted to these
    points.
    (b)~Layer polarization $|\zeta_\nu|$ of Eq.~(\ref{eq:zeta}) as a function of layer
    potential, at those hybridizations.  The $K$ valley saturates
    almost immediately while $\Gamma$ remains nearly unpolarized across the
    entire range.  The two vertical lines mark the interfacial potential of
    R-stacked bilayer WSe$_2$: the value inferred from the optical
    measurement of Ref.~\cite{Li2026_RWSe2} ($\varphi_0 = \UFE$~meV, dotted)
    and the value our own calculation gives ($U^K = \UFEdft$~meV, dash-dotted);
    Equation~(\ref{eq:threeU}) separates the three.  The dashed
    curve shows how a finite $t_K^{\rm R}$ would soften the $K$ response.
    In 2H spin--layer locking forbids the spin-conserving hopping and leaves
    the orbital matrix element finite; in R stacking spin permits the hopping
    and the three-fold rotation forbids the hopping at $K$
    itself~\cite{Akashi2015_3RMoS2}.%
  }
  \label{fig:calibration}
\end{figure}

We use our own converged density-functional-theory calculations for
bilayer WSe$_2$ (VASP~\cite{Kresse1996_vasp,Kresse1996_cms} with projector augmented-wave
potentials~\cite{Blochl1994_paw,KresseJoubert1999_paw}, the PBE
functional~\cite{Perdew1996_pbe} with spin--orbit coupling
(SOC), and DFT-D3
dispersion with Becke--Johnson
damping~\cite{Grimme2010_d3,Grimme2011_bj}).  The cutoff is 600~eV, the
relaxed lattice constant $a=\alatTwoH$~\AA{} and the interlayer W--W distance
$\dWW$~\AA.  $\Delta_{K\Gamma}$ was evaluated at
$E_z = 0, \pm 0.1, \pm 0.2, \pm 0.3, \pm 0.5$ and $\pm 1.0$~V\,nm$^{-1}$.
Figure~\ref{fig:calibration}(a) shows those values with
Eq.~(\ref{eq:dkgD}) through them, which has nothing left to adjust once
the splittings are read; Appendix~\ref{app:rparams} gives the settings.

\subsection{A layer potential acts as a valley potential}
\label{sec:outcome}

Figure~\ref{fig:calibration}(b) plots the layer-potential dependence of
the layer polarization $|\zeta_\nu|$ of the two 2H valleys.  At $\varphi_0 = \UFE$~meV, the potential an
\emph{R-stacked} polarization supplies~\cite{Li2026_RWSe2} and the first
of the two vertical lines, the $K$ valley is already fully polarized onto
one layer, $|\zeta_K| \to 1/2$, while $|\zeta_\Gamma|$ stays smaller by more
than an order of magnitude.  Both the band gap of a bilayer transition metal
dichalcogenide~\cite{Ramasubramaniam2011_stark,Zhang2019_WSe2efield} and
the interlayer excitons of bilayer
WSe$_2$~\cite{Huang2022_indirectX,Sevik2025_interlayerX} are long known to
respond to a perpendicular field.  The scan here gives a different
quantity, the separation between the two valence edges,
$\Delta_{K\Gamma}$.

Neither coupling comes from a fit.  At zero field the two layer states of
the $\Gamma$ doublet have equal on-site energies.  The zone-center
splitting is then exactly $2t_\Gamma$ and gives
$t_\Gamma^{\rm 2H} = \tGtwoHband$~meV directly (Appendix~\ref{app:bands}).
At $K$ the coupling vanishes by symmetry, for a different reason in each
stacking, and Sec.~\ref{sec:twoforbid} gives both reasons and the
calculations that check them.

The conclusions rest on the inequality
$t_\Gamma^{\rm 2H} \gg t_K^{\rm 2H}$, and we determine both couplings.  The scan resolves the two doublets separately, so
every parameter is read off one feature and $\Delta_{K\Gamma}(D)$ follows
with nothing left to adjust, to \residcanon~meV across the range.
Sec.~\ref{sec:counting} gives that accounting, and what a scan of
$\Delta_{K\Gamma}$ alone could not have supplied.

\subsection{\texorpdfstring{$t_K$}{tK} vanishes by spin--layer locking in
2H and by rotation in R}
\label{sec:twoforbid}

In 2H the two layers are antiparallel.  States at a given $K$ point in
opposite layers carry opposite spin, and interlayer tunneling is forbidden
in a spin-conserving description: $t_K^{\rm 2H} = 0$.  The zero-field bands give
$t_K^{\rm 2H} = 0$ to the numerical precision of the calculation,
$\tKtwoHbound$~meV (Appendix~\ref{app:rparams}), and we impose the
symmetry value throughout.

In R stacking the layers are parallel and the spins match,
and spin permits the tunneling [Fig.~\ref{fig:registries}(b)].  A three-fold rotation forbids
the tunneling instead, and more strongly.  Both polar registries keep that rotation.  Each layer has three
inequivalent three-fold axes: one through the center of a hexagon, one
through a chalcogen and one through the metal.  After the lateral shift, each axis of the
bilayer passes through one kind of site in the lower layer and another
kind in the upper.  Moving the rotation center between two such sites
translates a Bloch state by the vector $\mathbf d$ joining them, and
$\mathbf K\cdot\mathbf d = 2\pi/3$.  The $K$-point Bloch states of the two
layers are eigenstates of that rotation with eigenvalues
differing by a factor $e^{2\pi i/3}$.  Thus, the interlayer matrix
element between them vanishes at $K$.  Away from $K$ the little group is smaller and no such rule
applies.  Reference~\cite{Akashi2015_3RMoS2} proves this for 3R-MoS$_2$,
both by first-principles calculation and analytically.  The authors note
that the proof uses only the crystal and orbital symmetries and is therefore
guaranteed across the 3R-M$X_2$ family.

The two suppressions are of different kinds, and the R-stacked one is the
more robust.  In 2H the three pairings of axes give a
net phase of zero at the valence-band top~\cite{Akashi2015_3RMoS2}.  The
rotation permits the orbital matrix element, which is nonzero, and the
spin--orbit splitting $\lambda$ confines each state to one layer all the
same.
The confinement is measured by $\lambda/\sqrt{\lambda^2+t_b^2}$, in which
$t_b$ is that orbital amplitude and not the spin-conserving $t_K$ of
Eq.~(\ref{eq:Delta})~\cite{Akashi2015_3RMoS2}, and the ratio is close to one in a tungsten
compound~\cite{LeBoeuf2015_SOC}.  The size of $\lambda$ depends on the
atomic number of the metal.  In R the phases differ by a third of a turn, and the matrix element
vanishes whatever $\lambda$ is.  The lateral shift that creates the polarization
also switches the hopping off: stack the layers directly over one another
and the phases agree again.

Our own bands show the node in the interlayer coupling $t(\mathbf k)$
directly.  At the compensation field the two on-site energies coincide, $\Delta_K = 0$,
and the layer projection is well conditioned there.  The
coupling rises linearly from $K$, at
$\tslopeKM$ and $\tslopeKG$~meV\,\AA{} in the two inequivalent directions,
with intercepts of $\tinterKM$ and $\tinterKG$~meV.  That is a linear zero at $K$, not a small value [Eq.~(\ref{eq:tnode}), Appendix~\ref{app:bands}].
A coupling that merely happened to be small would stay small in a
neighborhood of $K$.

\section{The R-stacked landscape}
\label{sec:landscape}

In R stacking the polarization supplies a layer potential of its own, and
the curve $\Delta_{K\Gamma}(D)$ moves with it.  The polarization moves the minimum of that curve off
$D = 0$, to the field that cancels the polarization.  Reversing the
polarization sends the minimum to the opposite field, exactly, and the
distance is calculated.  Changing the stacking renormalizes the
curve $\Delta_{K\Gamma}(D)$ itself.  The shape transfers and the two energy
scales do not, and no symmetry constrains that part.  This section
determines the interlayer step and the $K$ hopping of the polar registries
from the bands, calibrates the field axis the displacement is measured on, and
maps $\Delta_{K\Gamma}$ over the strain--field plane.

\subsection{The polarization moves the minimum to the compensation field}

\begin{figure*}[tb]
  \includegraphics[width=0.95\textwidth]{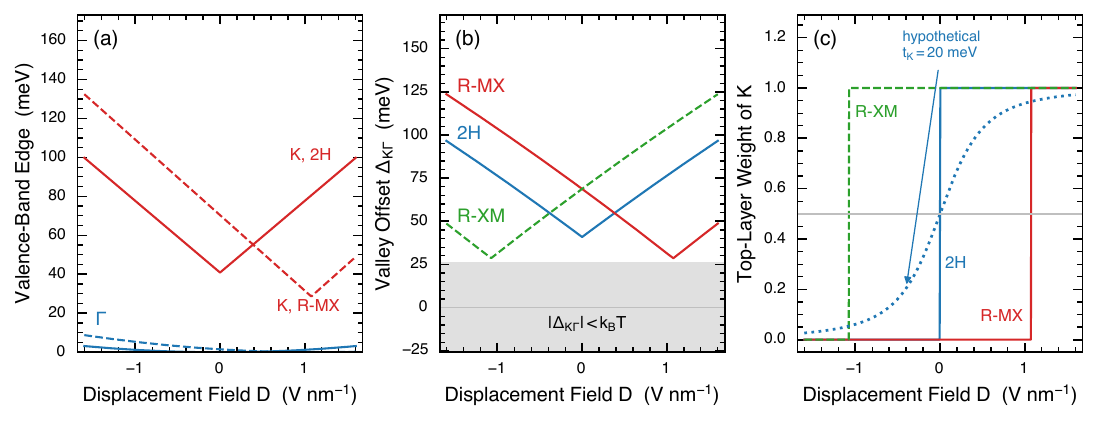}
  \caption{%
    Valley-selective Stark response.
    (a)~Displacement-field dependence of the $K$ and $\Gamma$
    valence-band edges, for 2H (solid) and R-MX (dashed).  Only the $K$ pair
    is named for both stackings: the two $\Gamma$ curves stay within
    $\GsepMax$~meV of one another across the whole range, which is the
    statement that the stacking barely moves $\Gamma$.
    (b)~The landscape $\Delta_{K\Gamma}(D)$.  The two R curves are computed
    from the directly determined R-stacked zero-field parameters of
    Sec.~\ref{sec:A1}, not by translating the 2H curve, and map into one
    another under $D \to -D$.  The field axis uses the dipole length
    $d_K = \dKR$~nm obtained from the R-stacked field scan itself
    (Sec.~\ref{sec:fieldaxis}).  The shaded band marks $|\Delta_{K\Gamma}| < k_BT$ at
    300~K.
    (c)~Top-layer weight of the $K$ band edge.  Each stacking switches
    layers at the field that cancels its own polar term, 2H at $D=0$ and the
    two R registries at $\pm D^{*}$.  In the $t_K\to 0$ limit that symmetry imposes, the step is
    abrupt; a finite hopping would round the corner.  The dotted curve is 2H with a
    hypothetical $t_K = 20$~meV.  2H and R-XM coincide at
    saturation, and both names appear there.%
  }
  \label{fig:stark}
\end{figure*}

Figure~\ref{fig:stark} shows where the field dependence of
$\Delta_{K\Gamma}$ comes from.  The curves use the R-stacked parameters that
Sec.~\ref{sec:A1} determines directly, and not the nonpolar ones
transferred.  The $K$ edge follows the
layer potential with slope $\pm e d_K/2$ and develops a cusp at the field that
compensates the built-in polarization, while the $\Gamma$ edge remains almost
flat [Fig.~\ref{fig:stark}(a)].  The landscape $\Delta_{K\Gamma}(D)$ is even
about $D=0$ for 2H and about $D = \pm D^{*}$ for R-MX and R-XM
[Fig.~\ref{fig:stark}(b)], the two R curves being mirror images as required by
Eq.~(\ref{eq:mirror}).  The $K$-edge layer weight switches between the two layers over a field range governed by $t_K$.
With $t_K$ at zero the switch is abrupt.  Each stacking switches at the field that
cancels its own polar term: 2H at $D = 0$, the two R registries at
$\pm D^{*}$ [Fig.~\ref{fig:stark}(c)].

R stacking moves the system \emph{further} into the $K$-dominant regime,
consistent with the $K$ valence-edge assignment of
Ref.~\cite{Li2026_RWSe2}, and moves the $K$/$\Gamma$ crossover further out
in tension.

\subsection{The polarization is converted into a valley shift at almost the
full ceiling}
\label{sec:A1}

The polar term of R stacking has two parameters.  The field scan returns a step of $U_K^{\rm eff} = \UFEdft$~meV and, without imposing the selection
rule, returns a central $t_K^{\rm R} = \tKR$~meV with
$t_K^{\rm R} < \tKlimitcanon$~meV.  The interlayer step shifts the $K$ edge by
\begin{equation}
\sqrt{(t_K^{\rm R})^2 + (U_{\rm FE}/2)^2} - t_K^{\rm R},
\label{eq:shiftvstk}
\end{equation}
which for $t_K = 0$ is half the step, $U_{\rm FE}/2 = \UFEhalfdft$~meV, and
that is the most any one valley can take from a step of \UFEdft~meV.  A
finite coupling falls short of that ceiling by less than $t_K^{\rm R}$
itself.  The limit above therefore places the $K$ edge within
\tKlimitcanon~meV of that ceiling. At zero field the two layer-derived subbands of a valley
are split by
$\delta E_\nu = 2\sqrt{t_\nu^2+(U^\nu/2)^2}$, and the upper branch has a
layer polarization
$\langle\tau_z\rangle_\nu = s\,(U^\nu/2)/\sqrt{t_\nu^2+(U^\nu/2)^2}$,
where $s$ indicates the registry, that is, the direction of the
polarization: the upper branch weights the bottom layer in R-MX and the top
layer in R-XM.  Since the polar step is positive in
Eq.~(\ref{eq:Delta}), the modulus appears when these are inverted
in closed form,
\begin{equation}
U^\nu = \delta E_\nu \left|\langle\tau_z\rangle_\nu\right|,
\qquad
t_\nu = \tfrac{1}{2}\,\delta E_\nu \sqrt{1-\langle\tau_z\rangle_\nu^2}.
\label{eq:zerofield}
\end{equation}
One calculation of one registry, reporting the splittings and the layer
weights, therefore gives $U^\nu$ with no field scan and no fitting.  The
same calculation gives a hopping that cannot be used, for the reason
Appendix~\ref{app:tkroute} explains, and the hopping quoted above comes
from the field scan.  The polarization-induced band shift follows without
even assuming a value for $U_{\rm FE}$, since $\sqrt{t_\nu^2+(U_{\rm FE}/2)^2}=\delta E_\nu/2$
identically:
\begin{equation}
\sqrt{t_\nu^2+(U_{\rm FE}/2)^2}-t_\nu
 = \tfrac{1}{2}\,\delta E_\nu\left(1-\sqrt{1-\langle\tau_z\rangle_\nu^2}\right).
\label{eq:shiftfree}
\end{equation}

We computed the two registries separately, each from its own
self-consistent run.  Their agreement is then a test of the calculation.
Appendix~\ref{app:rparams} gives the settings, which include the dipole
correction a polar slab needs~\cite{NeugebauerScheffler1992_dipole}, the
convention used for the layer weights, and four independent checks on the
inversion: the two spin--orbit multiplets against each other, the two
registries against each other, the vacuum-level step against the
band-derived potential, and the $\Gamma$ doublet.

Inverting Eq.~(\ref{eq:zerofield}) on each multiplet and averaging gives an
interfacial potential of \Uproj~meV, within \Uprojpct\% of the field-scan
value below.  The hopping does not survive the same inversion.
Equation~(\ref{eq:zerofield}) returns $t_\nu$ through
$\sqrt{1-\langle\tau_z\rangle_\nu^2}$, which is ill conditioned exactly where
a layer-decoupled valley sits, and a leakage of under two per cent in the
projection onto the atomic spheres accounts for the whole of the \tKproj~meV
returned at $K$ (Appendix~\ref{app:tkroute}).  The field scan needs no such
resolution and gives
\begin{equation*}
U_K^{\rm eff} = \UFEdft~\text{meV},
\qquad
t_K^{\rm R} = \tKR~\text{meV}
\;\;(<\tKlimitcanon~\text{meV}).
\end{equation*}
These are the values quoted above and used throughout.
The value of $U_{\rm FE}$ exceeds the optical one~\cite{Li2026_RWSe2} by
\UFEpct\%, and the vacuum-level step by \vacpct\%.  These are three different
quantities, and the next subsection separates them.

\label{sec:whichtK}%
With the directly computed offset $\Delta_{K\Gamma}^{\rm R}(0) = \dkgRdft$~meV,
these quantities define the R-stacked bilayer without reference to 2H, and
every R-stacked result below uses them.  Those results hold anywhere in the
range the bound on $t_K^{\rm R}$ allows, because the argument needs only the
contrast $t_\Gamma^{\rm R} \gg t_K^{\rm R}$ and the conversion efficiency
that contrast gives.

\subsubsection*{The field dependence of
\texorpdfstring{$\Delta_{K\Gamma}$}{Delta(K-Gamma)} follows from the
splitting, with nothing fitted}

The field scan determines the whole field dependence of $\Delta_{K\Gamma}$ with
no fitted parameter.  The $K$ edge rises by half the doublet splitting
while $\Gamma$ rises by the hybridized amount the same potential drives.
Thus,
\begin{equation}
\delta\Delta_{K\Gamma}
 = \tfrac{1}{2}\,\delta(\delta E_K)
 - \delta\!\left[\sqrt{t_\Gamma^2 + (\delta E_K/2)^2}\right],
\label{eq:dkgpredict}
\end{equation}
where $\delta$ denotes the change from zero field.  Every quantity on the
right is already fixed: $\delta E_K$ is computed at each field, and
$t_\Gamma^{\rm R}$ comes from the zero-field $\Gamma$ doublet.  The dipole
length does not enter, because the field appears only through the calculated
splitting.  Writing one layer potential for both valleys is an assumption of this
expression and not of the calibration, which measures the two separately;
the departure is \UratiopctG\% (Sec.~\ref{sec:counting}).  The residual below shows the error,
which is small because $t_\Gamma$ dominates the square root.
Writing that potential as $\delta E_K$ also takes $t_K^{\rm R}=0$, since the
two differ by $\sqrt{\Delta_K^2+4t_K^2}-|\Delta_K|$.  Even at the upper bound
of Eq.~(\ref{eq:tKfield}) that changes $\Delta_{K\Gamma}$ by under \dkgpredtK~meV.  Equation~(\ref{eq:dkgpredict}) reproduces the observed
$\Delta_{K\Gamma}$ across a $\pfrange$~meV swing with a root-mean-square
residual of $\pfresidxm$~meV for R-XM and $\pfresid$~meV for R-MX, the
largest single deviation being $\pfresidmax$~meV at
$\pfoutlier$~V\,\AA$^{-1}$.  The
scans give both columns to four decimals, so what is left is the model's
own error and not the precision of the input: the prediction writes one
layer potential for both valleys, and Sec.~\ref{sec:counting} measures the
two to differ by \UratiopctG\%.  The two-level description
holds for the relative motion of the valleys as well as for the splitting
of one of them.

\subsubsection*{Three quantities, not one}

Three quantities have been called the interfacial potential at various points,
and they differ.  Each needs a separate name.

\begin{equation}
\begin{aligned}
U_K^{\rm eff} &= \UFEdft~\text{meV},\\
\Delta\varphi_{\rm WW} &= \dphiRzero~\text{meV},\\
\varphi_0 &= \UFE~\text{meV},
\end{aligned}
\label{eq:threeU}
\end{equation}
where $U_K^{\rm eff}$ is the inversion of the $K$ doublet through
Eq.~(\ref{eq:zerofield}), $\Delta\varphi_{\rm WW}$ is the planar-averaged
electrostatic potential evaluated at the two W planes, and $\varphi_0$ is the
optical estimate of Ref.~\cite{Li2026_RWSe2}.  The three sample different
things, and Appendix~\ref{app:rparams} explains what each one samples and why
the band-derived step is the largest of them.  The field-induced changes of the band-derived and
electrostatic steps agree to the precision Appendix~\ref{app:rparams}
reports in Eq.~(\ref{eq:faithful}).

\label{sec:liscales}%
Below, $U_{\rm FE}$ means $U_K^{\rm eff} = \UFEdft$~meV, the step the $K$
valley sees, because the polar physics here rests on that valley.  The
$\Gamma$ valley sees a step \UratiopctG\% smaller and a dipole length
smaller in nearly the same proportion, and the model carries both; the
notation is shorthand and not an approximation.  One quantity notices the
difference between those two proportions, the zero of $g_d$, and
Sec.~\ref{sec:electrical} is where it is felt.  On the experimental side
Ref.~\cite{Li2026_RWSe2} reports a valence-band offset
$\Delta_{\rm v} = \DvOffset$~meV beside $\varphi_0$, and the two differ by
\liinconsist\%.  We propagate both experimental values through the results below
and take that \liinconsist\% as a lower bound on the systematic uncertainty
of the polar term, for the reason Appendix~\ref{app:robust} locates.

\subsubsection*{The landscape is displaced \emph{and} renormalized}

\emph{Polarization displaces the curve $\Delta_{K\Gamma}(D)$ along the field
axis, and changing the stacking renormalizes the same curve.}  Symmetry
protects the displacement and says nothing about the renormalization.  Thus,
only the first half of Eq.~(\ref{eq:shift}) holds.

The relation between the two \emph{registries} is exact.
Equation~(\ref{eq:mirror}) is a symmetry statement, independent of every
parameter: R-MX and R-XM are field-axis mirrors of one another, centered on
opposite compensation fields $D = -sD^{*}$.  Thus, the polarization determines \emph{where} the curve sits on the field
axis,
and reversing the polarization reverses the displacement.  Symmetry alone secures this, whatever the
calculation returns.

Stacking changes the two energy scales of that curve, and the R curve is
not a translated copy of the nonpolar one.  Going from 2H to R moves
$\bar E_K - \bar E_\Gamma$ and relaxes to a different geometry, and the best
rigid translate of the computed 2H curve still departs from the R one by up
to \rigidshift~meV.  The $\Gamma$ hybridization is nearly the same in both: read straight off
the zone-center splitting, $t_\Gamma$ is \tGtwoHband~meV in the
nonpolar bilayer and \tGRband~meV in R (Appendix~\ref{app:bands}).  Appendix~\ref{app:transfer}
gives the comparison, together with the transferability hypothesis the
comparison replaces: that the layer-symmetric parameters carry over from 2H
and the polarization enters only as the polar term $s\varphi_0$, the step taken from the optical experiment.  At zero field that
transferred model gives $\Delta_{K\Gamma}^{\rm R} = \dkgR$~meV
where the direct calculation gives \dkgRdft~meV.  The strain axis repeats
that comparison.  Both crossovers divide an offset
by the same computed slope, and the transferred model's
$\epscrossRtrans\%$ biaxial tension therefore stands to the calibrated
model's $\epscrossR\%$ in the ratio of the two offsets.  Section~\ref{sec:strain}
computes the crossover directly and finds $\straincrossRown\%$ on the
lattice constant the model uses, which is the calibrated value.  The shape of the field dependence transfers and the two energy scales,
horizontal and vertical, do not.  Formally,
\begin{equation}
\begin{aligned}
\Delta_{K\Gamma}^{\rm MX}(D) &= \Delta_{K\Gamma}^{\rm XM}(-D)
  \quad\text{exactly,}\\[2pt]
\Delta_{K\Gamma}^{\rm R}(D) &\neq \Delta_{K\Gamma}^{\rm 2H}(D + sD^{*}).
\end{aligned}
\label{eq:newstory}
\end{equation}

\subsection{The field axis}
\label{sec:fieldaxis}

The dipole length $d_\nu$ converts an applied field into an interlayer
energy, $e d_\nu D$, and therefore determines the scale of the field axis.
Rescaling that axis would not change the couplings, the selection rule, the
registry mirror or the zero-field inversion, which are zero-field
quantities, ratios of them, or relations that hold at every field.
Rescaling would change the compensation field $D^{*}$, the Stark slopes, the
magnitude of $g_d$, the read-out contrast and the screening ratio, all of
which enter through $d_\nu$.  Two independent calculations give the length.  The field
scan of the R-stacked registries returns $d_K = \dKR$~nm from the Stark
slope of the $K$ doublet.
The nonpolar scan of Sec.~\ref{sec:calibration} returns $\dKtwoH$~nm from
the nonpolar $K$ splitting, a different observable on a different
structure.  The two agree to $\dagreepct\%$, and they share no input.  The
polar length corresponds to $\varepsilon_\perp = \epsR$, taking the
geometric W--W distance of that registry, $\dWWR$~\AA.

The planar-averaged potential confirms that length with no band structure
at all.  At $0.5$~V\,nm$^{-1}$ the induced potential difference
between the two tungsten planes is $\dphitwoHfree$~meV in the nonpolar
bilayer and $\dphiR$~meV in R-MX.  The two registries agree to
$\registryagree$~meV and give $\varepsilon_\perp = \epsTwoHfree$ and
$\epsRfree$.

Published values for this material are lower: $\epsperpWSe$ from the
linear-response calculations of Ref.~\cite{Laturia2018_dielectric} and
$\epsperpYuan$ from the field-dependent band structures of
Ref.~\cite{Yuan2013_zeeman}, both for the nonpolar form.  The five
determinations collected here span $\epsperpYuan$ to $\epsRfree$, and the
dipole length used throughout is the one our own three agree on.

The two hybrid values, $\epsperpHSEtwoH$ for the nonpolar bilayer and
$\epsperpHSE$ for the polar registry, both lie below the $\epsperpWSe$
reported for this material, and the two semilocal values lie above that
figure, so the published value falls between the functionals and not between the
stackings.  Raising the functional therefore moves the screening towards
the published value in both stackings, which is the strongest external
check we have on the field axis.  The splitting stays
exactly linear in the field at both levels, and each functional gives a
dipole length with nothing fitted; Appendix~\ref{app:rparams} gives the
matched-pair protocol and the individual values.

Every finite-field result here is either of a polar cell or was computed
with the charge-density symmetrization disabled, which a centrosymmetric
cell in a field requires: left on, the symmetrization removes the screening
response entirely (Appendix~\ref{app:rparams}).

\subsection{Strain--field phase diagram}
\label{sec:strain}

Strain is quantitatively usable as a tuning variable even where the
absolute offset stays uncertain.  The slope survives the
exchange--correlation functional far better than the offset does.  On two
series that share every structure and differ only in the functional, the
slope goes from \strainslopePBE~meV per percent at the semilocal level to
\strainslopeHSE~meV per percent with the hybrid, a change of
\strainslopefuncpct\%, where the same change of functional more than
doubles $\Delta_{K\Gamma}$ itself.  The crossover moves with the offset
and not with the slope, from $\epscrossTwoH\%$ to $\epscrossHSE\%$, and
the hybrid series brackets that crossover with calculated points, as the
semilocal one does.  Both series run over the full $\pm2\%$ on the
relaxed geometries.

\begin{figure*}[tb]
  \includegraphics[width=0.98\textwidth]{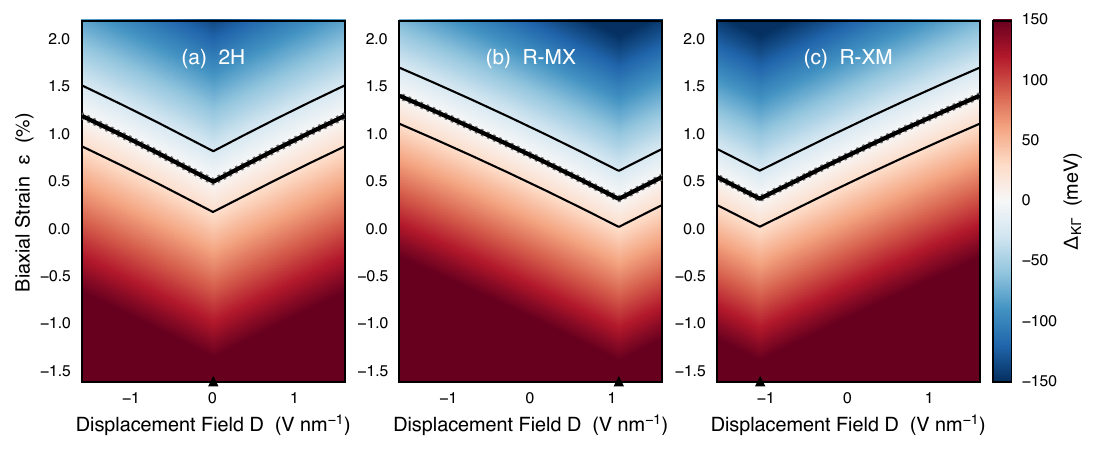}
  \caption{%
    Strain--field valley phase diagram for (a)~2H, (b)~R-MX and (c)~R-XM,
    each from the parameters calculated for that stacking
    (Secs.~\ref{sec:A1}--\ref{sec:strain}).  Color encodes
    $\Delta_{K\Gamma}$, red for a $K$ edge and blue for a $\Gamma$ edge.
    The scale is diverging, and in a grayscale reproduction the two signs have the same tone and the heavy contour is what separates them:
    $K$ lies below the contour and $\Gamma$ above.  That contour is
    $\Delta_{K\Gamma}=0$, and the
    thin and gray dotted pairs bound $|\Delta_{K\Gamma}| < k_BT$ at 300 and
    at 25~K, the 25~K pair tracking the zero contour closely.  The marker on
    each field axis is $-sD^*$, the vertex of that panel's zero contour.
    The panels are the calibrated model.  The direct scans of
    Sec.~\ref{sec:strain} locate the crossovers a few hundredths of a
    percent above these in all three panels.  Those scans are referred to the
    nonpolar lattice constant (Appendix~\ref{app:rparams}), and on the polar
    registry's own constant the directly calculated crossover is
    $\straincrossRown\%$, which is the value the model gives.%
  }
  \label{fig:phase}
\end{figure*}

Figure~\ref{fig:phase} maps $\Delta_{K\Gamma}$ over the strain--field plane for
the three stackings.  The
$K$-dominant, mixed and $\Gamma$-dominant regions are delimited by
$|\Delta_{K\Gamma}| = k_BT$ at 300~K and at 25~K, the latter relevant to the
low-temperature optical measurements.  The R-MX and R-XM panels, Figs.~\ref{fig:phase}(b) and~\ref{fig:phase}(c),
map into one another under $D \to -D$, which is exact.  Each of them uses
the R-stacked parameters of Sec.~\ref{sec:A1}, and departs from a rigid
translate of panel~(a) by up to \rigidshift~meV.

About one percent of biaxial tension reaches the $K$--$\Gamma$ crossover in
R-stacked WSe$_2$.  A biaxial-strain scan of the polar bilayer at zero
field gives $\Delta_{K\Gamma}$ falling linearly at $\strainslopeR$~meV per
percent, to $\strainresid$~meV over $\pm2\%$, and changing sign at $\straincrossR\%$ tension, between the calculated points at $+0.75\%$ and $+1\%$.  $\Delta_{K\Gamma}$ falls from $\dkgRstrainzero$~meV
at zero strain to $\dkgRone$~meV at $+1\%$ and $\dkgRtwo$~meV at $+2\%$, where
$\Gamma$ becomes the valence-band maximum.  This series is referred to the
nonpolar lattice constant, as both are, which puts the two slopes on one
axis.  The polar registry's own $a_0$ is the larger of the two, and the
unstrained end of the series therefore sits above the $\dkgRdft$~meV of
Sec.~\ref{sec:A1}.  Substrate mismatch, flexible supports and
process-compatible stressors deliver strain of this order~\cite{He2013,Conley2013,Desai2014_WSe2strain,Carrascoso2021_strainTMD,Kumar2024_strainFET,Jaikissoon2024_strainCMOS,Ahn2017NatComm},
but the largest biaxial tension transferred into a WSe$_2$ transistor channel
is \strainxfer\%~\cite{Michail2024_straintransfer}, so the crossover lies
above what a device process has demonstrated.  First-principles work on
biaxially strained monolayers finds the same sign: tension lowers the valence
band at $K$ and leaves $\Gamma$ untouched, which moves the valence-band
maximum from $K$ to $\Gamma$~\cite{Johari2012}.  Biaxial
tension of this size is also reported to raise the electron mobility of monolayer
dichalcogenides~\cite{Yang2024_biaxialstrain}, which is a different carrier in a
different system from the holes considered here.  Our 2H data place the $K$--$\Gamma$ crossover at $\epscrossTwoH\%$.

The $\Gamma$ mass responds to strain far more strongly than the $K$ mass.
Between $-2\%$ and $+2\%$, $m^*_\Gamma$ changes from $\mGstrainlo$ to
$\mGstrainhi$~$m_0$ while $m^*_K$ changes only from $\mKstrainlo$ to
$\mKstrainhi$~$m_0$.  The response is not linear, so we quote the
fractional sensitivity: $\mGstrainpct\%$ per percent of strain at $\Gamma$
and $\mKstrainpct\%$ at $K$, a ratio of $\massratioStrain$.  These
masses sit on the same structures as the slope above, and the zero-strain
point of the series returns the $m^*_K$ and $m^*_\Gamma$ of
Sec.~\ref{sec:A1} to better than $2\%$.  This is consistent with $t_\Gamma \gg t_K$, since the valley built from
interlayer-hybridized orbitals is the valley that responds to a change of
geometry.
However, the two statements remain distinct.
The strain response of an effective mass also collects changes in the
in-plane hopping, the crystal field, mixing with other bands, and whatever
out-of-plane relaxation accompanies the in-plane strain.  We read the agreement as
corroboration, not as a second measurement of the hybridization.  At zero strain the two masses are $\mKR$ and
$\mGR$~$m_0$; the 2H reference gives $\mstarK$ and $\mstarG$.  The two
stackings therefore agree on the $K$ mass, and come close on the heavier
$\Gamma$ mass, as they do on the strain slope.

The nonpolar slope quoted above, $\strainslopePBE$~meV per percent,
differs from the R value by $\strainslopeagree\%$, both scans running over
$\pm2\%$ on one footing.  The two stackings therefore respond to strain
to within that \strainslopeagree\%, and the separation between their
crossovers comes almost entirely from $\Delta_{K\Gamma}(0)$.

The diagram is where the conflicting $K$- and $\Gamma$-edge assignments can be
placed.  Both lie inside the plotted range, separated by of order one percent
of biaxial strain at fixed field, or equivalently by a fraction of the
interlayer-distance variation that our own van der Waals comparison spans.  We
present this as a plausible reconciliation, not as an assignment: the
experiments differ in substrate, screening environment and strain state, and
distinguishing among those causes requires the local strain
characterization discussed in Sec.~\ref{sec:discussion}.

\section{Layer-resolved valley thermodynamics}
\label{sec:thermo}

Putting carriers on the two valleys turns the landscape of
Sec.~\ref{sec:landscape} into an occupation, and the count extends over the layer-resolved subbands and not over the valleys alone.  The $K$ valley of an R-stacked bilayer has two subbands split by
$\delta E_K = \dEKMXband$~meV, which is \Kdoubletsplit~$k_BT$ at room
temperature.
Both are occupied, and both shift with the field.  This section counts
them, and two susceptibilities follow: the response of the
valley occupation to the carrier density, and the response to the
displacement field.  A single gate reaches the field but locks field and density together at $D = qp/2\varepsilon_0$, and the field response is a dual-gate quantity (Appendix~\ref{app:gates}).

\subsection{The \texorpdfstring{$K$}{K} doublet is thermally active}
\label{sec:doublet}

\begin{figure*}[tb]
  \includegraphics[width=0.95\textwidth]{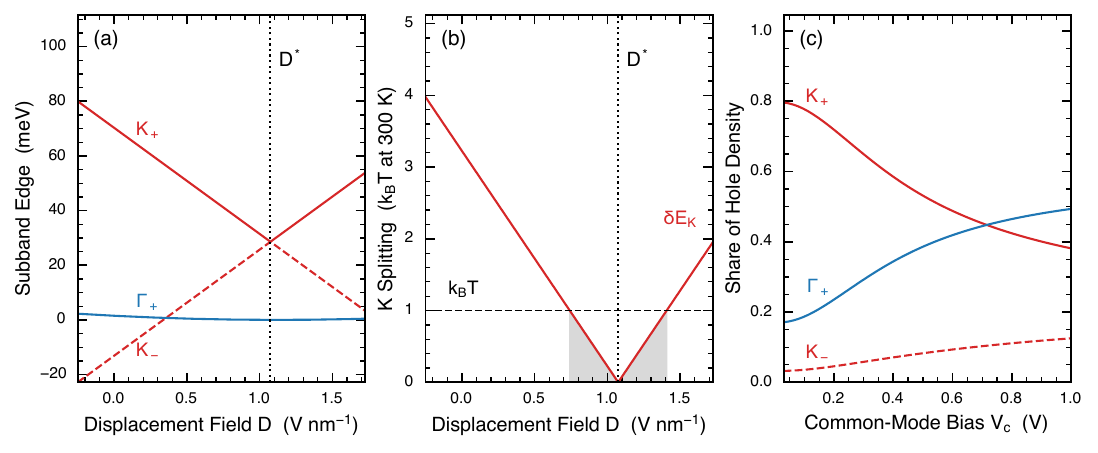}
  \caption{%
    The $K$ layer doublet is thermally active, for R-MX at 300~K.
    (a)~Displacement-field dependence of the subband edges.  The two $K$ branches
    approach most closely at the compensation field $D^{*}$, where the applied
    field cancels the built-in polarization.  The $\Gamma$ edge stays nearly
    flat.
    (b)~$K$ doublet splitting in units of $k_BT$, on the same axis.  The
    closing of the doublet is resolved only in panel~(b).  A finite
    $t_K^{\rm R}$ would open the minimum to $2t_K^{\rm R}$.  The three-fold
    selection rule requires a closed minimum and the scan returns one, so
    the curve is drawn closed.  The bound of
    Eq.~(\ref{eq:tKfield}) would open the minimum to $\dEKminbound$~meV, and
    the band calculation at the compensation field gives $\dEKDstar$~meV
    (Appendix~\ref{app:bands}).  Both openings are invisible against the $100$~meV span of (a).  The shaded region marks where the
    splitting falls below $k_BT$ and the $K$-edge density of states is
    effectively doubled; the region spans $\kTwindow$~V\,nm$^{-1}$ about $D^{*}$.
    (c)~Common-mode-bias dependence of the share of the total hole
    density held by each subband.  The lower $K$ branch is never negligible and more than
    doubles that share across the sweep; the lower $\Gamma$ branch, at
    $\GdoubletkT\,k_BT$, is omitted because the share is of order
    $\Gbranchshare$.%
  }
  \label{fig:subbands}
\end{figure*}

Each valley contributes two layer-derived subbands, and for the $K$ valley
both must be retained.  The two are split by
$2\sqrt{t_K^2+(\Delta_K/2)^2}$, which for R stacking at zero displacement
field is $\delta E_K = \dEKMXband$~meV, or $\Kdoubletsplit\,k_BT$ at 300~K
[Fig.~\ref{fig:subbands}(a)].  The splitting is close to the interfacial potential
$U_{\rm FE} = \UFEdft$~meV the field scan returns, as must be the case, since the
two coincide for a layer-decoupled valley, and the three-fold selection
rule forces $t_K^{\rm R}(K)$ to zero.  The scan bears that out, closing
the splitting to \dEKnode~meV at the compensation field, below the
\dEKtwoHzero~meV the nonpolar stacking returns for a doublet symmetry makes
exactly degenerate.  What separates the two figures quoted here is not a
hopping but the band step: \dEKMXband~meV comes from a
non-self-consistent path and \UFEdft~meV from the self-consistent scan,
and the \bandscf~meV between them is the same difference every entry of
the atlas carries (Appendix~\ref{app:bands}).
The $\Gamma$ doublet, split by
$2\sqrt{t_\Gamma^2 + (\Delta_\Gamma/2)^2} = \dEGzero$~meV
($\GdoubletkT\,k_BT$), is by contrast entirely inert: the lower branch accounts for a fraction $\Gbranchshare$ of the holes and the problem is effectively a
three-subband one.

Resolving the layers also determines the density of states unambiguously, and the
ratio that enters is the degeneracy-weighted one.
Following Ref.~\cite{PaperI_valleyGAA} the 2D prefactor
includes an explicit valley degeneracy,
$N_{2D,\nu} = g_\nu m^*_\nu m_0 k_BT/2\pi\hbar^2$, with $g_K = 4$ for the 2H
bilayer --- a doublet degenerate under the combined parity--time
($\mathcal{PT}$) operation at each of the two
pockets $K$ and $K'$ --- and $g_\Gamma = 2$ from spin alone.  The ratio that
enters the statistics is therefore
\begin{equation}
r_N = \frac{g_\Gamma m^*_\Gamma}{g_K m^*_K} = \rN,
\label{eq:rN}
\end{equation}
half the bare mass ratio $m^*_\Gamma/m^*_K = \rmass$.  Those are the 2H
masses, which define the convention; on the R-stacked masses of
Sec.~\ref{sec:A1} that definition gives $r_N = \rNR$, and every R-stacked
result below uses that value.

The layer resolution makes this simple, because in 2H the
$\mathcal{PT}$ doublet at $K$ \emph{is} the layer doublet.  That doublet is
observable: Fourier analysis of Shubnikov--de Haas
oscillations in a gated 2H bilayer resolves the two subbands as localized on
the top and on the bottom layer~\cite{Fallahazad2016_mass}.  Once the two
branches are kept explicitly, the degeneracy left for each is two --- the
$K,K'$ multiplicity at $K$, and spin at $\Gamma$ --- so every branch has
the full $m^*_\nu m_0 k_BT/\pi\hbar^2$.  Summing over branches reproduces
Eq.~(\ref{eq:rN}): at zero field the two $K$ branches are degenerate and give
$2m^*_K$, while at $\Gamma$ only the upper branch is thermally active and
gives $m^*_\Gamma$.  In the Boltzmann limit this yields
$f_\Gamma = \fGpaperone$ at the nonpolar $\Delta_{K\Gamma}=\dkgTwoHPBE$~meV and
300~K,
against which the implementation is checked directly.

The number of thermally occupied subbands changes as the gate
is swept, and the occupations below follow continuously, with no band
counting.  Occupations follow from Fermi--Dirac statistics applied to all four
subbands independently.  Thus, the lower $K$ branch fills smoothly as the
common mode raises the chemical potential, and contains \KlowerA\% of the holes at a common-mode gate bias $V_c = 0.05$~V and \KlowerB\% at $V_c = 1.0$~V [Fig.~\ref{fig:subbands}(c)].  Because each subband
contributes a step to the 2D density of states, that staircase also enters
the quantum capacitance~\cite{Luryi1988_QC,Bennett2023_QC} and hence
both transconductances automatically.

The second branch has real weight.  Removing the lower $K$
branch, as a single-band-per-valley treatment would, changes $f_\Gamma$ by
\dropfGpct\% and $|g_d|$ by \dropgdpct\%.  The removal leaves the antisymmetry
between the registries untouched, because that follows from
Eq.~(\ref{eq:mirror}) and holds for any truncation of the subband list.  The layer resolution determines the magnitudes and leaves untouched the
relation between the registries, which is the division drawn throughout.

Sweeping the differential mode drives the $K$ splitting through a
\emph{minimum} at the compensation field $D = -sD^{*}$, where the applied
field cancels the built-in polarization exactly.  This follows from
$\Delta_K^{(s)}(D) = sU_{\rm FE} + ed_KD$.  The two polar registries give one curve,
displaced to opposite fields [Fig.~\ref{fig:registries}(c)]; whether the
nonpolar reference belongs to that family is a separate question, taken up
in Appendix~\ref{app:transfer}.  In a generic two-level system a finite hybridization would open that minimum to $2t_K^{\rm R}$ and make the
crossing an avoided one.  However, the three-fold selection rule of
Sec.~\ref{sec:calibration} forces $t_K^{\rm R}(K)$ to zero.  Thus, the crossing
is symmetry-protected, and a calculation placed at the compensation field itself
(Appendix~\ref{app:bands}) bottoms at $\dEKDstar$~meV, below the \dEKtwoHband~meV that band
step returns for a doublet that symmetry makes exactly degenerate.  The $K$-edge density of states is
therefore nearly doubled there, and stays so across the $\kTwindow$~V\,nm$^{-1}$
about $D^{*}$ over which the splitting remains below $k_BT$, falling back
outside that window [Fig.~\ref{fig:subbands}(b)].  That doubling partly offsets the simultaneous minimum of
$\Delta_{K\Gamma}$, and the $f_\Gamma$ maximum in
Fig.~\ref{fig:thermo}(c) is broader than the landscape alone would suggest.

\subsection{Statistics and dual-gate electrostatics}

Populations follow from 2D Fermi--Dirac statistics,
$p_b = N_{2D,b}\ln(1+e^{\eta_b})$, where
$N_{2D} = m^*_\nu m_0 k_BT/\pi\hbar^2$ per branch, in the counting of
Eq.~(\ref{eq:rN}), and $\eta_b$ is the Fermi level measured from
the edge of subband $b$ in units of $k_BT$, evaluated for all four subbands, with the
layer-resolved densities weighted by $w_{\rm top}$ and $1-w_{\rm top}$.  The
total equilibrium density is held fixed and the flat-band Fermi level solved
accordingly.  In the limit of vanishing layer splitting the four-subband count reduces
exactly to the single-band-per-valley description of
Ref.~\cite{PaperII_transconductance}, reproducing the Boltzmann result
$f_\Gamma = r_{\rm DOS}e^{-x}/(1+r_{\rm DOS}e^{-x})$ with
$x=\Delta_{K\Gamma}/k_BT$.

Throughout we use the degeneracy-weighted ratio $r_N$ of
Eq.~(\ref{eq:rN}), with all four subbands counted explicitly.  Refs.~\cite{PaperI_valleyGAA} and~\cite{PaperII_transconductance}
differ on that convention.  The choice changes the numerical value of $\chi_v^{(n)}$ and leaves
everything else here intact.  Appendix~\ref{app:robust} states which
convention each paper uses, which anyone comparing susceptibilities across
the three will need.

In a dual-gated device [Fig.~\ref{fig:registries}(d)] the top and bottom
gate voltages $V_t$ and $V_b$
separate into a common mode $V_c = (V_t+V_b)/2$ and a differential mode
$V_d = (V_t-V_b)/2$, controlling
respectively the total density and the displacement field.  Treating the
bilayer as a sheet at surface potential $\psi_s$ between capacitances $C_t$
and $C_b$,
\begin{equation*}
C_t(V_t-\psi_s) + C_b(V_b-\psi_s) = q\,p + C_{it}\psi_s ,
\end{equation*}
in which $C_{it}$ is the interface-trap capacitance, and
which for symmetric gates reduces to the single-gate relation of
Ref.~\cite{PaperII_transconductance} with $C_{ox}\to C_t+C_b$.  The
difference of the same two gate terms is the displacement field,
\begin{equation}
D = \frac{C_t(V_t-\psi_s) - C_b(V_b-\psi_s)}{2\varepsilon_0}.
\label{eq:Dgates}
\end{equation}
For symmetric gates the surface potential cancels from
Eq.~(\ref{eq:Dgates}) identically, and with $C_{ox} =
\varepsilon_0\varepsilon_{ox}/t_{ox}$, in which $\varepsilon_{ox}$ and
$t_{ox}$ are the relative permittivity and the thickness of each gate
dielectric, the field follows from the gate difference alone,
\begin{equation}
  D = \frac{\varepsilon_{ox}}{2\,t_{ox}}\,(V_t - V_b)
    = \frac{\varepsilon_{ox}}{t_{ox}}\,V_d ,
  \label{eq:Dvd}
\end{equation}
so the abscissa of every figure here is a differential gate sweep.
The conversion brings in the gate dielectric.  Thus, we report in $D$ and
give the gate voltages that reach a given $D$ alongside.  The compensation field $D^{*}$ is
reached at $V_d = \Vdstar$~V on the scaled stack and at
$V_d \approx \VdstarhBN$~V on 20~nm hBN --- a factor of thirty between two
ordinary devices, for one and the same physical field.
Two susceptibilities follow, an occupation response and a landscape response,
\begin{equation}
\chi_v^{(n)} = \left(\frac{\partial f_\Gamma}{\partial V_c}\right)_{\!D},
\qquad
\chi_v^{(D)} = \left(\frac{\partial f_\Gamma}{\partial D}\right)_{\!n}.
\label{eq:chis}
\end{equation}
The field acts on the $K$ doublet twice over.  Thus, the landscape response
has two terms,
\begin{equation}
\chi_v^{(D)}
 = \frac{\partial f_\Gamma}{\partial \Delta_{K\Gamma}}
   \frac{\partial \Delta_{K\Gamma}}{\partial D}
 + \frac{\partial f_\Gamma}{\partial \delta E_K}
   \frac{\partial \delta E_K}{\partial D}.
\label{eq:chainrule}
\end{equation}
Moving towards $-sD^{*}$ lowers $\Delta_{K\Gamma}$, which fills
$\Gamma$, while at the same time closing $\delta E_K$, which populates the lower $K$ branch and retains holes at $K$.  The two work against each other.
At $V_c = 0.30$~V the $\delta E_K$ term removes \chiKcancelzero\% of the
$\Delta_{K\Gamma}$ term at $D = 0$ and \chiKcancelhi\% at
$0.8$~V\,nm$^{-1}$.  There is no third term
from $\Gamma$: the $\Gamma$ doublet is split by \dEGzero~meV, far beyond reach at any temperature considered here.  The lower branch stays empty, and $f_\Gamma$ is independent of
$\delta E_\Gamma$.  We evaluate
Eq.~(\ref{eq:chainrule}) as the derivative of the self-consistent solution at
fixed density.

The second susceptibility has no analogue in a single-gated device, because
one gate cannot move the landscape independently of the occupation.  The
state of the channel depends on $(p,D)$ and on nothing else about the gates.  The three gate architectures are three regions of one plane: a wrapped
gate reaches the line $D = 0$, a single gate the line
$D = qp/2\varepsilon_0$, and only a dual gate the interior.  The results
here therefore specialize to the architectures of Refs.~\cite{PaperI_valleyGAA}
and~\cite{PaperII_transconductance} by restriction to their lines, and Appendix~\ref{app:gates} places the three side by side with the fields each
can reach.

\begin{figure*}[tb]
  \includegraphics[width=0.95\textwidth]{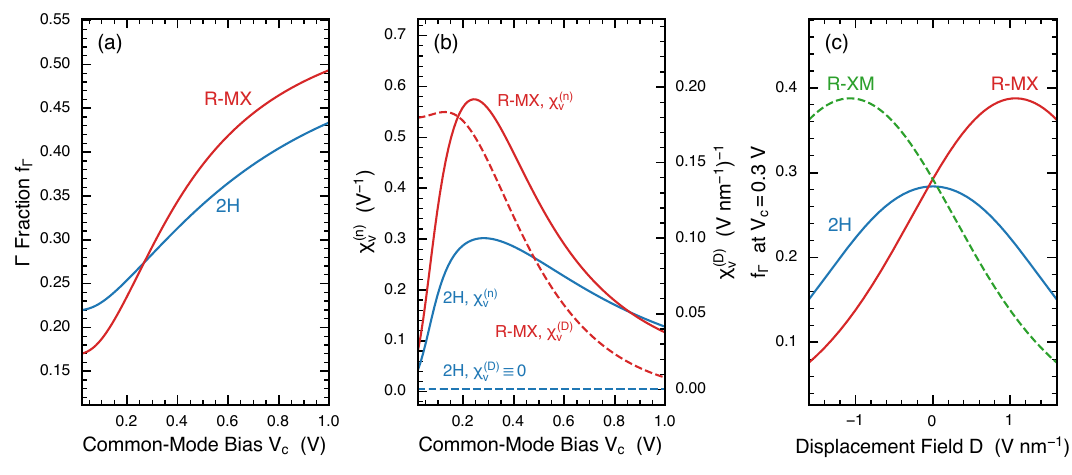}
  \caption{%
    Layer-resolved valley thermodynamics at 300~K.
    (a)~Common-mode-bias dependence of the $\Gamma$-valley fraction for 2H and R-MX.
    (b)~The occupation susceptibility $\chi_v^{(n)}$ (left axis, solid) and the
    landscape susceptibility $\chi_v^{(D)}$ (right axis, dashed) of
    Eq.~(\ref{eq:chis}), for both stackings; color denotes the stacking as in
    (a).  The nonpolar bilayer has $\chi_v^{(D)} \equiv 0$ at $D = 0$, which is
    Eq.~(\ref{eq:parity2H}) and not a numerical accident: $\Delta_{K\Gamma}$ and $\delta E_K$ are both even in $D$ there, so both terms of
    Eq.~(\ref{eq:chainrule}) vanish and the landscape is stationary under a
    differential bias.  Panel~(b) is the one place in which the two stackings
    differ in kind and not merely in degree.
    (c)~Displacement-field dependence of $f_\Gamma$ at fixed common mode.  The
    maximum sits at $D=0$ for the nonpolar reference and at $D = \pm D^{*}$
    for R-MX and R-XM.%
  }
  \label{fig:thermo}
\end{figure*}

The peak of the $\Gamma$ fraction moves with the polarization, and that
displacement is the thermodynamic fingerprint of the stored polarization.
Figure~\ref{fig:thermo} shows the common-mode-bias dependence of the
$\Gamma$ fraction in panel~(a), the two susceptibilities in panel~(b), and $f_\Gamma$
across the field axis in panel~(c).  Along the field axis the $\Gamma$ fraction peaks where $\Delta_{K\Gamma}$ is smallest, at
$D=0$ for 2H and at $D=\pm D^{*}$ for the two polar registries.

\section{Common- and differential-mode transport}
\label{sec:transport}

The common mode cannot see the polarization and the differential mode can,
which makes the stored polarization an electrical observable.

In the linear regime the drain current is $I_D = (W/L)\,\mu_{\rm eff}\,q\,p\,V_{DS}$, with $W/L$ the channel
width-to-length ratio, $q$ the elementary charge, $p$ the mobile hole sheet
density and $V_{DS}$ the drain bias.  The effective mobility
$\mu_{\rm eff} = \mu_K - (\mu_K-\mu_\Gamma) f_\Gamma$ weights the
single-valley hole mobilities $\mu_K$ and $\mu_\Gamma$ by the occupation,
following Ref.~\cite{PaperII_transconductance}.  Only mobile charge contributes, trapped
charge being electrostatically active but immobile.  We take
$\mu_K/\mu_\Gamma = m^*_\Gamma/m^*_K = \rdosR$ for R stacking, that is, equal
intravalley scattering times; the masses are the R-stacked ones of
Sec.~\ref{sec:A1}, $\mKR$ and $\mGR\,m_0$, and not the 2H values
quoted in Sec.~\ref{sec:intro}.  The ratio is a modeling choice and not a computed transport
result~\cite{Jin2014,Sohier2019_valleyEng,Afrid2026_intervalley,Ha2025_WSe2mobility}, and
Appendix~\ref{app:robust} scans the ratio.  Measured hole mobilities in mono- and bilayer WSe$_2$ span a
wide range and depend strongly on contacts and dielectric
environment~\cite{Allain2014_WSe2mobility,Movva2015_WSe2dualgate},
with quantum-oscillation measurements reaching the degenerate regime this
model treats~\cite{Xu2017_QHE}.  The two transconductances are
$g_c = \partial I_D/\partial V_c$ and $g_d = \partial I_D/\partial V_d$.

\begin{figure*}[tb]
  \includegraphics[width=0.95\textwidth]{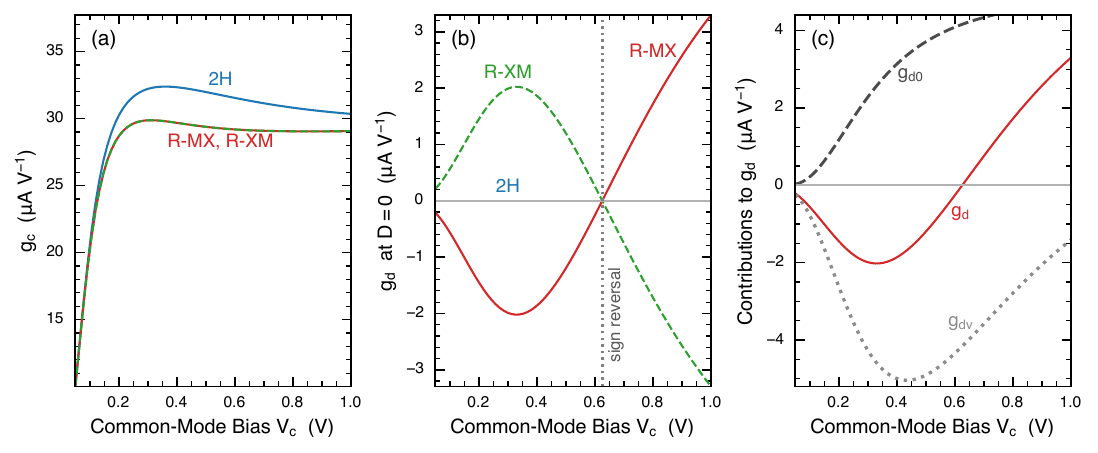}
  \caption{%
    The two transconductances at 300~K.
    (a)~$g_c$ is identical for the two polar registries and blind to the
    polarization sign.  The nonpolar reference differs by at most \gcpct\%.
    (b)~$g_d$ at zero displacement field vanishes for 2H in the ideal
    top--bottom-symmetric geometry assumed here and reverses with the
    registry.  The dotted line marks the common-mode bias at which the two
    contributions in panel~(c) cancel.  Symmetry guarantees the antisymmetry
    between the registries and not the size of either: at that bias both registries vanish together, and the polar response is finite only away from that bias.
    (c)~Decomposition of $g_d$ for R-MX into the density term $g_{d0}$ and the
    landscape term $g_{dv}$, which have opposite signs.%
  }
  \label{fig:gm}
\end{figure*}

The common-mode response [Fig.~\ref{fig:gm}(a)] is identical for the
two registries and within \gcpct\% of the nonpolar reference.  That response measures accumulation and occupation
redistribution, and is blind to the sign of the stored polarization.  The differential-mode response
[Fig.~\ref{fig:gm}(b)] behaves differently in kind.  For the nonpolar
reference $g_d$ vanishes identically at $D=0$, because $\Delta_{K\Gamma}(D)$
is even and the derivative odd; this is an exact consequence of
inversion symmetry, under the conditions listed below, and a useful control.  For R stacking the two
registries give equal and opposite $g_d$ at $D=0$, finite except at one
common-mode bias where both vanish together.

Two different symmetry statements are in play here, and they are easily
confused.  The mirror relation Eq.~(\ref{eq:mirror}) relates the two
registries to \emph{each other},
\begin{equation}
I_{\rm MX}(D) = I_{\rm XM}(-D)
\;\;\Longrightarrow\;\;
I'_{\rm MX}(0) = -\,I'_{\rm XM}(0),
\label{eq:parityR}
\end{equation}
so at $D=0$ the currents coincide and the direct-current contrast vanishes
identically.  The two slopes at $D=0$, however, are equal and opposite.

Both statements are exact only under the conditions in which they were
derived, and those are worth listing.  They are a centrosymmetric 2H
reference, a top--bottom symmetric environment so that $U_{\rm env}=0$ in
Eq.~(\ref{eq:Delta}), a dual-gate stack symmetric enough that the common and
differential modes separate cleanly, a single equilibrium domain, and
transport in which the contacts and the scattering rates are themselves
field-symmetric.  The first three are design
requirements on the measurement and are discussed in
Sec.~\ref{sec:symmetry}.  The last two are idealizations of the model.
Symmetry protects one statement: no parameter value can make the 2H slope
finite.  A real device may still miss the zero, and neither slope is
required to vanish.

The nonpolar reference is the special case in which the
mirror maps the system to \emph{itself},
\begin{equation}
I_{\rm 2H}(D) = I_{\rm 2H}(-D)
\;\;\Longrightarrow\;\;
I'_{\rm 2H}(0) = 0,
\label{eq:parity2H}
\end{equation}
and under the conditions just listed $g_d(D=0)$ vanishes for the nonpolar
reference at every operating point, which no parameter value can change.  The stored
polarization is thus read out at zero displacement field through the sign of a
small-signal differential-mode response --- experimentally, a lock-in
measurement with a small a.c.\ differential bias --- even though no direct
current contrast exists there.  A finite read field remains an alternative:
at $V_d = \Vdread$~mV on the scaled stack, which reaches
$D = \Dread$~V\,nm$^{-1}$, the direct-current contrast is \readoutdft\%.

$g_d$ is the sum of two competing terms, and near the assumed mobility
ratio they nearly cancel [Fig.~\ref{fig:gm}(c)].  For R-MX, $D=0$ lies on the flank descending to the compensation field, and raising the differential bias lowers $\Delta_{K\Gamma}$ and fills the heavy $\Gamma$ valley.  The added density of
states raises the carrier density at fixed common mode, a positive term
$g_{d0}$, while the transfer out of the light, high-mobility $K$ valley gives
a negative one, $g_{dv}$.  Their near cancellation makes $g_d$ change sign
as a function of common-mode bias.  The magnitude is correspondingly sensitive to
$\mu_K/\mu_\Gamma$.  The antisymmetry between registries, by contrast, follows
from Eq.~(\ref{eq:mirror}) through Eq.~(\ref{eq:parityR}), under the
conditions listed with that equation, and is exact for any mobility ratio.  Any
experimental protocol should rely on the sign reversal at $D=0$,
with the common mode at a fixed bias away from the cancellation, and not on the absolute magnitude.

\subsection{The differential mode is where the stackings separate}
\label{sec:whygd}

How far 2H and R stacking differ depends on which channel is measured, and
the answer tells an experiment where to look.  The dual-gate geometry assumed throughout is
standard for 2D channels~\cite{Radisavljevic2012_dualgate} and has
been used to address valley populations
electrically~\cite{Lai2023_valleytronic}.

In direct-current transport they are barely distinguishable.  At $V_c=0.3$~V
and $D=0$, R-MX has a drain current differing from the 2H value by
only \dIDpct\%,
even though the R-MX landscape sits \dkgdiffdft~meV higher.  The reason is a
cancellation between two large effects, which are worth separating.

Going from 2H to R changes the valley occupation in two competing ways: the
landscape rises and the $K$ layer doublet splits.  The two nearly cancel,
and $f_\Gamma$ changes by only $\dfGpct$\%.  Appendix~\ref{app:robust} separates
them.  The layer doublet is not a correction to the landscape effect but a
comparable and opposing one, which is the degeneracy bookkeeping of
Sec.~\ref{sec:doublet} seen from the other side.  Still less reaches
the current, where a mobility change and a density change cancel again.
Scanning $\Delta_{K\Gamma}(0)$ across the entire van
der Waals range of Sec.~\ref{sec:calibration} never makes the current
difference exceed about \dIDscanmax\%.  A comparison of $I_D$--$V_G$
characteristics between 2H and 3R regions is a weak test of the
mechanism, and weak for a structural reason.

In the differential channel they differ in kind.  At $D=0$ the two $K$ branches of 2H are degenerate.  They are equally occupied, and the holes divide evenly between the layers, $f_{\rm top} =
\ftopTwoH$.  Each branch is itself layer-polarized, with opposite spin on
opposite layers.  However, the pair has no net imbalance.  In R the
polarization splits them, and the occupied edge sits on one layer or the
other according to the registry, $f_{\rm top} = \ftopR$.
Correspondingly $g_d(D=0)$ is exactly zero for 2H, by the parity argument
above and under the conditions stated there, and $\pm\gdRdft$~$\mu$A\,V$^{-1}$ for the two R registries.  At this
operating point the contrast is between zero and finite, a difference in
kind.  Symmetry protects the equal-and-opposite relation between the
polar registries.  Symmetry leaves the size of either open.

The direct-current contrast and the
landscape effect move oppositely as $t_K^{\rm R}$ grows.  Thus, a large
measured contrast would indicate a \emph{weak} landscape effect.  The two together
are an independent check on the computed $t_K^{\rm R}$, which Appendix~\ref{app:robust} gives.

\subsection{Where to read: degeneracy screens the differential response}
\label{sec:readpoint}

One further consequence of the layer resolution bears directly on how a
read-out should be biased, and is not obvious from the landscape alone.

The differential-mode valley response is a purely non-degenerate effect and
is screened away as the channel fills.  In two dimensions the density of states is constant, giving $p \propto \ln(1+e^{\eta})$, and when the
$K$ doublet splits by $\pm u$ in units of $k_BT$ the two branches
contribute $\ln(1+e^{\eta-u}) + \ln(1+e^{\eta+u})$.  In the degenerate limit
$\eta \gg u$ each logarithm becomes linear, the sum tends to
$2\eta$, and the splitting cancels exactly.

Quantitatively, at $u = 1$ the change in $p_\Gamma/p_K$ approaches the
non-degenerate value $1/\cosh(1) - 1 = -\screenNonDeg\%$ at low density but
falls to $-\screenDegen\%$ once $\eta \simeq 1$.  On the scaled stack that condition is
reached by $V_c \simeq 0.3$~V, and the direct-current read-out contrast at
$V_d = \Vdread$~mV is \readoutdft\% there and \readoutLowdft\% at
$V_c = 0.05$~V, where the density is $\pLow$~cm$^{-2}$ in place of
$\pHi$~cm$^{-2}$.

Ferroelectric read-out should therefore be performed at low carrier
density, near subthreshold.  The mechanism itself remains --- the registry
antisymmetry of $g_d$ holds at any density, though the magnitude that
antisymmetry applies to is screened away in the degenerate limit.  However, the operating point does depend on screening, and the prescription is opposite to
the usual instinct of biasing hard to maximize the current.

Biasing the device hard anyway gives up more than signal strength.  Strong
accumulation reverses the direct-current contrast, so an experiment run
there can assign the polarization backwards, and
Appendix~\ref{app:robust} gives the crossing density, the fractional
response and the dependence on the mobility ratio.  The zero of $g_d$
meanwhile remains at $-sD^{*}$ at every density we have examined, because
the zero is a statement about parity and not about occupation.  That is
the clearest reason to use the locator of Sec.~\ref{sec:electrical} and
not a read-out based on amplitude.

The compensation field is within reach of an ordinary gate stack.  The
scaled stack of
Refs.~\cite{PaperI_valleyGAA,PaperII_transconductance}
(equivalent oxide thickness $\mathrm{EOT}=\EOT$~nm,
$C_{ox}\approx\Cox$~mF\,m$^{-2}$) delivers
$\Dstack$~V\,nm$^{-1}$ per volt of differential bias.  The compensation
field is reached at $V_d = \Vdstar$~V, and the whole ferroelectric landscape
is traversed within a few hundred millivolts.  A conventional
hBN-encapsulated dual gate needs $V_d \approx \VdstarhBN$~V for the same
field, inside the $\pm\gateLi$~V that Ref.~\cite{Li2026_RWSe2} sweeps on
each gate.  A gate-all-around geometry cannot reach that field at
all, for the structural reason given in Table~\ref{tab:gates}.

Two statements are symmetry protected, under the conditions listed above: $g_d$ vanishes at $D=0$ for the nonpolar reference, and the two polar registries
give equal and opposite values there.  Both follow from
Eq.~(\ref{eq:mirror}) and hold for any functional, any mobility ratio and
any operating point.  Model dependent are the magnitude $|g_d|$, the
direct-current contrast, and the common-mode bias at which the density and
landscape terms cancel, all three through $f_\Gamma$ and the last two
through the assumed mobility ratio.  That $g_d$ is nonzero at all belongs
with them: equal and opposite permits both to vanish, and at the
cancellation bias they do, so the read-out is operated away from a bias
whose position is model dependent.

The experiment this suggests is accordingly a reversal, with no absolute
reading required.  Holding the operating point fixed and rewriting the polarization
should change the sign of the differential-mode response, with the magnitude
returning to the same value; the nonpolar reference should show no response at
$D=0$ in either state.  Framed that way the measurement tests the symmetry
statement directly and is insensitive to everything in the model-dependent
list, since the device, the contacts and the bias point are common to the two
readings.

\subsection{A thermal route to the compensation field}
\label{sec:arrhenius}

The parameters above are also measurable, in one device and with no
first-principles input.  In the non-degenerate limit the $K$ layer doublet is
the only source of curvature in an Arrhenius plot of the valley occupation,
and that curvature is minimized at $D = -sD^{*}$, where the applied field
cancels the polar term, whatever the parameters are.  The minimum therefore
sits at opposite fields for the two registries, and locating the minimum
tests Eq.~(\ref{eq:mirror}) in one device and with no 2H reference sample,
which makes the route a test of the central claim and not of one parameter.
Appendix~\ref{app:arrsys} gives the relation, what the position and the depth
can each be asked for, and the limits of the method.

\subsection{An electrical route to the compensation field}
\label{sec:electrical}

A parity argument places a zero of the differential-mode transconductance
at the compensation field, which makes that field readable at one
temperature and with nothing fitted.  The thermal route of
Sec.~\ref{sec:arrhenius} needs a temperature series.  This one does not.

At a displacement field that cancels both valleys' polar terms at once, every
$\Delta_\nu$ vanishes.  Within the two-level description of
Eq.~(\ref{eq:eigen}) each band edge depends on $\Delta_\nu$ only through $\Delta_\nu^2$.  Every band energy --- and hence the valley occupation, the
carrier density, the effective mobility and the drain current --- is an
\emph{even} function of $D$ about that field.  An even function has zero derivative at the center.  Thus, \begin{equation}
  \left.\frac{\partial I_D}{\partial D}\right|_{V_c,\,T} = 0
  \qquad\text{at}\qquad D = -sD^{*} ,
  \label{eq:gdnull}
\end{equation}

\noindent
at any temperature, at any operating density, and for any $K$-to-$\Gamma$
mobility ratio --- though not for any \emph{layer}-dependent one.  Three
conditions do enter, and each of the three matters: a symmetric gate stack,
a single field that cancels both valleys, and transport that cannot tell the
two layers.  The rest of this section takes them one by one.

\begin{figure*}[tb]
  \centering
  \includegraphics[width=\textwidth]{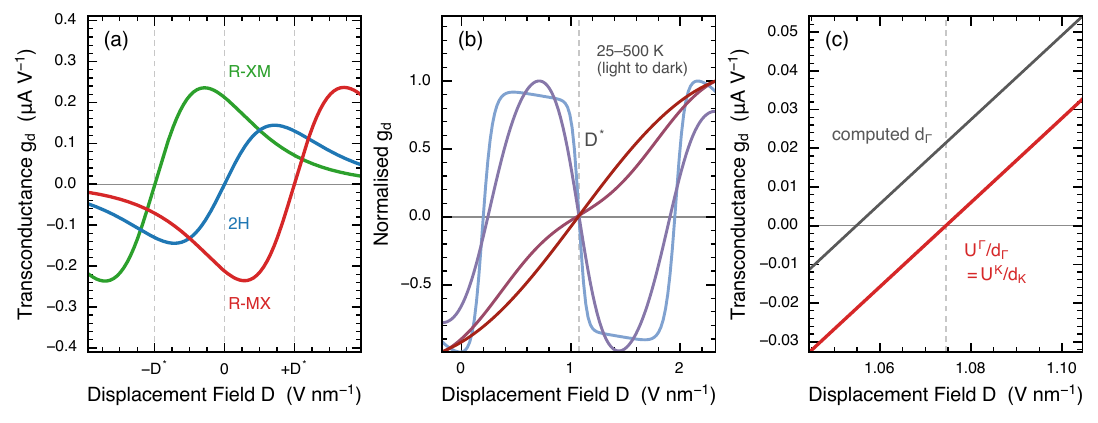}
  \caption{%
    \textbf{The compensation field read electrically.}
    Panels~(a) and~(b) are drawn in the limit in which the two valleys
    share one compensation field, $U^K/d_K = U^\Gamma/d_\Gamma$, where
    Eq.~(\ref{eq:gdnull}) is exact; panel~(c) uses the computed dipole
    lengths and steps, which differ, and shows how far the zero moves.
    (a)~Displacement-field dependence of the differential
    transconductance for the three stackings, at $300$~K and at the
    subthreshold bias $V_c = 0.05$~V, where the differential readout is
    strongest.  Each curve crosses zero once, at the field that cancels the
    polar term of that registry: $D=0$ for 2H, $+D^{*}$ for
    R-MX and $-D^{*}$ for R-XM.  Dashed lines mark those three fields.  All
    three fall back towards zero at large field, where $\Gamma$ is empty and
    the landscape has no carriers left to move.  At the $V_c = 0.30$~V of
    panels~(b) and~(c) the nonpolar curve crosses zero twice more, between
    $D=0$ and $\pm D^{*}$, where the loss of carriers cancels the gain in
    mobility; that pair is a cancellation and not a parity, and the
    zero at $D=0$ is where it is at every bias.
    (b)~The same quantity for R-MX, normalized to the maximum, at
    \gdTlo, $100$, $300$ and \gdThi~K (light to dark).  The shape is strongly
    temperature dependent and the zero is not, because
    Eq.~(\ref{eq:gdnull}) is a parity statement and not a thermal one.  The
    outer zeros that appear at low temperature are a mirror pair about the
    central one.
    (c)~What the two valleys' compensation fields differing by \nullgap\%
    does to the zero, on an expanded scale.  The gray curve carries the
    computed $\Gamma$-valley dipole length and step of Eq.~(\ref{eq:dratio}), and
    crosses zero \gdnullpct\% below $D^{*}$ (dashed).  The red curve is the
    same model with the $\Gamma$ step moved onto the $K$ valley's
    compensation field, so that one field compensates both valleys and the
    zero sits at $D^{*}$ exactly.%
  }
  \label{fig:gdnull}
\end{figure*}

On a symmetric dual gate the gates themselves do not drive the common mode as $V_d$ changes $D$, and the measured $g_d$ inherits the zero directly.  Equation~(\ref{eq:gdnull}) becomes a statement about $g_d$ itself.  That needs care, because the sweep leaves the carrier density free.  Sweeping $V_d$ at fixed $V_c$ moves
$\Delta_{K\Gamma}$, which redistributes the valleys and changes the total density of states, and $p$ drifts by a few percent across the range of
Fig.~\ref{fig:gdnull}(a).  However, the self-consistent solution
--- surface potential, carrier density, valley occupation and effective mobility alike --- is even about $-sD^{*}$.  The drift is even too, and the first derivative vanishes there.  We verify all four numerically, to machine
precision.  This is the generalization to the polar registries of the
nonpolar parity result $g_d(D=0)=0$ of Eq.~(\ref{eq:parity2H}); 2H is the
case $s=0$.

The zero sits where symmetry locates it and stays there
[Fig.~\ref{fig:gdnull}].  Panel~(a) shows the three stackings on one field
axis.  The zero sits at $D=0$ for 2H and at $\pm D^{*}$ for the two polar registries, and the \emph{position} of a zero encodes the stored polarization, while the size of the signal does not.  Panel~(b) varies the temperature
from \gdTlo{} to \gdThi{}~K.  The shape changes qualitatively below
\gdturnover~K, where $g_d$ reverses and acquires two further zeros, themselves
a mirror pair about $-sD^{*}$.  The central zero does not move.  The
zero is insensitive to carrier density and to $t_K^{\rm R}$ over the
whole admitted range, as Eq.~(\ref{eq:gdnull}) requires, since the $K$ term
is even in $\Delta_K$ whatever $t_K$ is.

The measurement is a small-signal one at a fixed operating point, and asks
for a low-resistance contact and a channel that is not contact-limited;
both have been demonstrated for WSe$_2$~\cite{Shen2021_contact,Pack2024_WSe2}.

Equation~(\ref{eq:gdnull}) is exact when one field cancels the polar term
in both valleys at once, $U^K/d_K = U^\Gamma/d_\Gamma$.  The two ratios are
computed, \dratioGK{} and \UratioGK, and the two valleys are therefore
compensated at \nullK{} and \nullG~V\,nm$^{-1}$, which lie \nullgap\%
apart.  The zero
of $g_d$ then sits \gdnullpct\% from $D^{*}$ at the higher density and
\gdnulllow\% at the lower one, at \gdnullD~V\,nm$^{-1}$ for a
$D^{*}$ of $\Dstardft$~V\,nm$^{-1}$ [Fig.~\ref{fig:gdnull}(c)].

The mismatch is \nullgap\% and the displacement far smaller, and the
displacement grows with density, for one reason: the $\Gamma$ valley holds
\fGRPBE{} of the holes at the higher of these two operating points and
less at the lower, so the $\Gamma$ compensation field has that much
leverage on the current and no more.  An experiment should therefore read
the zero at low density, which is the conclusion the degeneracy screening
of Sec.~\ref{sec:readpoint} reaches from the amplitude.

The device displaces the zero further than the bands do.  An asymmetric gate
stack removes the zero outright, layer-dependent transport displaces the zero
by \mulayershift\% for a \mulayer\% mobility asymmetry, a built-in
top--bottom step displaces the zero by \uenvshift\% for \uenvmev~meV, and a
spread in the local polarization leaves the zero at the device average;
Appendix~\ref{app:robust} treats each of them, together with the contact
resistance, which costs amplitude and leaves the zero where it is.  The
electrical route is therefore accurate to the degree that the device is
symmetric, and not to the degree that the band structure is: a percent needs
a percent-level device.

\section{Discussion}
\label{sec:discussion}

A dual-gated R-stacked WSe$_2$ device should be biased so that $n$ and $D$
are separated, and $g_c$ and $g_d$ should be measured independently.  The control is $g_d(D=0)=0$ for an identified 2H region.
Coexisting 2H and 3R regions in chemical-vapour-deposited bilayers supply that control on the same flake~\cite{DiBerardino2025_2H3R}, and the two stackings of bilayer WSe$_2$ are told apart optically~\cite{LiZ2022_stackingexcitons}.  The sign of $g_d$
should then track the polarization state written by a preceding pulse, and
should reverse when the domain is switched.  Because local strain of order one
percent moves the system across the phase diagram of
Fig.~\ref{fig:phase}, strain characterization is a prerequisite for
interpreting any $K$/$\Gamma$ assignment.

Three sources of uncertainty bear on the magnitudes, and none on the
symmetry statements of Sec.~\ref{sec:symmetry}.  The largest is the
dispersion treatment.  Across the three corrections we ran as a pair, so
that both stackings share every setting, the absolute offset of the polar
bilayer spans \vdwRlo{} to \vdwRhi~meV, enough to change which valley is
the band edge.  The difference between the stackings, the quantity this
paper claims, stays between \vdwdifflo{} and \vdwdiffhi~meV across the same
three.  Next is the exchange--correlation functional, which raises the
offset by $\dkgshiftR$~meV and the room-temperature occupation by a factor
of $\fGratio$, so predictions resting on the absolute occupation come as a
PBE--HSE06 band, with the R-stacked landscape underneath the PBE one.
Smallest, and no longer worth a band, is $t_K^{\rm R}$, which the canonical
scan bounds at \tKlimitcanon~meV.  That leaves the polarization-induced
conversion at the ceiling to the digits printed here.  The polarization-induced shift is far more robust than the offset: the absolute
offset does not enter Eq.~(\ref{eq:shiftvstk}) at all, and the functional
changes the shift by only \UFEHSEpct\%.  We present that shift as the
primary claim and the absolute landscape as a scanned axis.
Appendix~\ref{app:robust} gives the scans behind all three.

Everything here is equilibrium, single-domain physics.  Polarization reversal in rhombohedral TMDCs
proceeds through the release of pre-existing domain walls from their pinning
centers, which then propagate~\cite{Yang2024_domainwall}.  Coercive fields and switching times
therefore lie outside what this work can say.  A partially switched sample
would show the two registries averaged with the weight of each domain
fraction.  With the responses equal and opposite, a majority fraction $f$
of one registry gives $(2f-1)$ times the single-domain value: the magnitude
falls as the sample is switched, passes through zero at equal fractions, and
reverses once the opposite registry becomes the majority.  The measurement
therefore reads the domain balance, and not merely the presence of one state.

This work complements earlier
$\bm{k}\cdot\bm{p}$ modeling of R-stacked
bilayers~\cite{Lin2019_rashbaFE}, which addresses Rashba spin splitting
and how that splitting reverses with the polarization.  The present concern is the
$K$--$\Gamma$ occupation balance and the signature in transport.

\section{Conclusion}

We have established two results, on different kinds of evidence, and
together they specify one experiment.

\emph{By direct calculation.}  The interlayer hybridizations of the $K$ and
$\Gamma$ valence states of bilayer WSe$_2$ are of different kinds, not
different sizes.  A fit that assumes nothing about symmetry puts the $K$
coupling below \tKlimitcanon~meV, at the resolution of the calculation,
while the $\Gamma$ coupling is \tGRband~meV; a calculation at the field
that cancels the polarization finds the splitting there consistent with zero.
An interfacial polarization shifts the two valleys by different
amounts.  In R stacking
the band-effective layer asymmetry of the $K$ doublet is
$U_K^{\rm eff} = \UFEdft$~meV, and the nearly layer-decoupled $K$ edge
converts that step into a valley shift of \polshiftdft~meV, the ceiling
$U_K^{\rm eff}/2$.  $\Gamma$ barely shifts.  The induced interlayer potential drops of the two stackings agree to
$\registryagree$~meV, and the dipole length of the polar registry,
$\dKR$~nm, scales the field axis.

\emph{By symmetry, for any parameter values.}  Interlayer tunneling at $K$ is
strongly suppressed in the nonpolar reference by spin--layer locking, and
forbidden outright in the polar registries by the three-fold rotation.  The separation quoted above rests on a selection rule in the polar case, not on a coupling that merely happens to be small.  The two polar registries are
mirror images in the displacement field, $\Delta_{K\Gamma}^{\rm MX}(D) =
\Delta_{K\Gamma}^{\rm XM}(-D)$.  Whatever displacement the polarization
produces along the field axis, the registry reverses that displacement.  How far the landscape
displaces, $D^{*} = \Dstardft$~V\,nm$^{-1}$, is a calculated quantity and
belongs with the results above.  In transport the same relation makes the differential-mode
transconductance vanish at $D=0$ in an ideal top--bottom-symmetric nonpolar
reference, and be equal and opposite at $D=0$ between the two polar
registries.  Neither
depends on the functional, the mobility model or the operating point.  That
the polar response is nonzero does: equal and opposite permits both to
vanish, and at one common-mode bias they do.  The read-out is operated away
from that bias.  Magnitudes are model dependent throughout.

\emph{As an experimental test.}  The statement to measure is a reversal
and not a current difference: at $D=0$, with the common mode held at a
fixed bias away from the one at which the two contributions cancel,
rewriting the polarization should reverse the sign of the differential-mode
response, while the nonpolar reference shows none.  Sweeping the field, without rewriting the polarization, reads that parity a
second way, and a cheaper one.  The differential response passes through zero within about a percent
of the field that cancels the polarization, at a single temperature and with
no size needing calibration, and at opposite fields for the two registries
(Sec.~\ref{sec:electrical}).  The band structure contributes that percent; how near a measurement comes
depends on the symmetry of the device.  Bilayer WSe$_2$ has been grown
at scale~\cite{Chou2025_bilayerWSe2growth,Huang2014_WSe2CVDgrowth} and made
into p-type transistors operating at an estimated hole density of
$\pGhosh\times10^{13}$~cm$^{-2}$, above the top of the window used here~\cite{Ghosh2025_bilayerWSe2FET}, and p-type transistors on the R-stacked bilayer itself have already been reported~\cite{LiX2023_3RbilayerFET}; ferroelectric control of valley
populations has been demonstrated in a related
system~\cite{Li2022_ferroelectric_valley}, and ferroelectric oxides have been
integrated with 2D channels~\cite{Muller2011APL,Lin2025NatElectron}.  This work supplies the quantitative target: a landscape that moves by \polshiftdft~meV with the
polarization and a compensation field at $\Dstardft$~V\,nm$^{-1}$, which a
scaled dual gate reaches at $V_d = \Vdstar$~V.

\section*{Data availability}
The data that support the findings of this study are available from the
corresponding author upon reasonable request.

\begin{acknowledgments}
This work was supported by JSPS KAKENHI (Grants No.~JP26H02222,
No.~JP25K01609, No.~JP22H05473, and No.~JP21H01019).  The numerical calculations were
carried out on the Numerical Materials Simulator at the National Institute
for Materials Science.
\end{acknowledgments}

\appendix

\section{Computational settings and numerical checks}
\label{app:rparams}

The splitting of a valley doublet follows from the potential difference
between the layers, and that difference can be had without any band energy:
averaging the electrostatic potential over planes normal to the stacking
axis and taking the values at the two tungsten planes gives
$\Delta\varphi_{\rm WW}$.  At one applied field, in both stackings,
\begin{equation}
\begin{aligned}
\text{2H:}\quad & \delta(\Delta\varphi_{\rm WW}) = \dphitwoHfree~\text{meV},
  & \delta(\delta E_K) &= \dekIsymhalf~\text{meV},\\
\text{R-MX:}\quad & \delta(\Delta\varphi_{\rm WW}) = \dphiR~\text{meV},
  & \delta(\delta E_K) &= \dekRhalf~\text{meV}.
\end{aligned}
\label{eq:faithful}
\end{equation}
Each row compares two independent evaluations of one induced quantity, the
left from the planar-averaged potential and the right from the band
energies.  In R-MX both come from one run, and they agree to the
\bandprec~meV precision of the eigenvalues.  The 2H pair is drawn from
two calculations, the potential from the symmetrization controls of
Sec.~\ref{sec:fieldaxis} and the splitting from the canonical scan, and
agrees to $\faithfulpct\%$.  Thus, the bands follow the local potential,
and a dipole length can be read off either side.

The test is a stronger check than the zero-field inversion, because the test
lies far outside the calibration regime, and because the left-hand side
contains no band information.  The scope is narrow.  The
test says nothing about the center of either doublet, nor about the relative
motion of the two valley edges that $\Delta_{K\Gamma}$ measures.

\subsection{What the resolved scans determine, and what a scan of
$\Delta_{K\Gamma}$ alone could not}
\label{sec:counting}

Each parameter in Sec.~\ref{sec:outcome} is read off one feature of one
scan.  At zero field inversion symmetry makes the two layer states of the
nonpolar $\Gamma$ doublet equal in on-site energy, so the zone-center
splitting is exactly $2t_\Gamma$.  The $K$ doublet of the same calculation
must be degenerate for the same reason, and whatever that calculation
returns there measures the calculation and not the crystal.  Away from zero field each doublet widens as
$\delta E_\nu = 2\sqrt{t_\nu^2 + (\Delta_\nu/2)^2}$, which is exactly
quadratic in $\Delta_\nu$, so $\delta E_\nu^2$ is a quadratic polynomial in
the applied field and inverting it is linear least squares.  No starting
guess, no bounds and no optimizer enter, and $t_\nu$ appears only through
the constant term $U_\nu^2 + 4t_\nu^2$.  A hopping consistent with zero
therefore turns that term negative, and in the polar registry it does, by
\fourtKsq~meV$^2$.

With every parameter fixed this way, $\Delta_{K\Gamma}(D)$ is a prediction
with nothing left to adjust, and holds to \residcanon~meV across
$\pm1$~V\,nm$^{-1}$ in the nonpolar stacking and \residcanonR~meV over the
\nfieldsR{} fields of the polar scan.

A scan of $\Delta_{K\Gamma}$ alone could not have supplied this.
Equation~(\ref{eq:dkgD}) is even in $D$, so the fields at $-|D|$ repeat the
fields at $+|D|$, and two combinations stay invisible however many fields
are added, the blindness being a property of the response and not of the
sampling.  One is $t_K$ nearly
alone.  The other is the pair $(t_\Gamma, d_\Gamma)$, which enters only
through $d_\Gamma^2/t_\Gamma$, since $ed_\Gamma D/2$ stays well below
$t_\Gamma$ across the scan and the $\Gamma$ term is quadratic.  Scaling
$t_\Gamma$ by a quarter and moving $d_\Gamma$ to hold that combination
fixed moves the $\Gamma$ doublet by tens of meV, which the resolved scans
measure to \residGcanon~meV.

Resolving the doublets also tests $d_K = d_\Gamma$, which such a scan
could not.  The two are not equal.  The nonpolar scan gives
$d_\Gamma/d_K = \dratioGKtwoH$ and the polar one \dratioGK, so the two
valleys of either stacking sample an applied field in different proportions,
and imposing the equality raises the nonpolar residual from \residcanon{}
to \ansatzresid~meV.  The inequality has a plain reading.  The $\Gamma$
state is spread across the van der Waals gap while the $K$ state sits on one
layer, so a field moves the $\Gamma$ edge through a smaller effective
dipole.  The same asymmetry appears in the polar terms,
$U^\Gamma/U^K = \UratioGK$, and Sec.~\ref{sec:electrical} is where the
small difference between those two fractions is felt.

The resolution floor comes from the same campaign, and it is what the
bounds on $t_K$ rest on.  In the nonpolar stacking the $K$ doublet is
degenerate by symmetry and the two W planes are at equal potential, so the
\dEKtwoHzero~meV the calculation returns for the splitting, and the
\dphitwoHzero~meV between the two planes, are that calculation's error and not a
physical scale.  A doublet cannot split by less than $2t_K$.  The polar
registry closes the same doublet to \dEKnode~meV at the compensation field,
below that floor, so $t_K^{\rm R} \le \tKbound$~meV states what the
calculation can resolve and not what the crystal does.  The selection rule
of Sec.~\ref{sec:twoforbid} is what says the hopping vanishes.

We give the R-stacked settings in full, since several of the conclusions
above depend on them.  The slab is polar, and a dipole
correction along the stacking
axis~\cite{NeugebauerScheffler1992_dipole} is required: without the correction the zero-field $K$
splitting is \dEKnodipole~meV, while the corrected calculation on that
structure gives \dEKaug~meV, and every R-stacked value here comes from the
corrected calculations.  By the same
argument we disable the charge-density symmetrization wherever a field is
applied to a centrosymmetric cell, since inversion would otherwise forbid a
net dipole in the induced charge and remove the screening response.  The
result is insensitive to the smearing width over
$\sigma = 0.02$--$0.05$~eV.
  We relaxed the two stackings at their separate in-plane lattice constants, with the
ionic positions free, the cell height held fixed, and a force tolerance of
$5\times10^{-4}$~eV\,\AA$^{-1}$, on a $\Gamma$-centered
$24\times24\times1$ mesh~\cite{MonkhorstPack1976}, which contains $K$
exactly.  The two lattice constants differ: $a = \alatR$~\AA{} for the
polar registries and $\alatTwoH$~\AA{} for the nonpolar one, fixed as the
minimum of the energy curve in $a$ for each stacking.  A relaxation that stops on the
force criterion leaves the stress unexamined, so we took the minimum
instead of relaxing the cell.  The field scans hold the geometry fixed, so that the field and the
structure are not varied together.  The polar scan reads
$\Delta_{K\Gamma}$ on the same $24\times24$ mesh; the nonpolar one halves
the mesh to $12\times12$, because disabling the symmetrization there leaves
$144$ points where the symmetrized calculation has $19$, and the sweep
below puts that halving at \convmesh~meV.  The plane-wave cutoff is 600~eV, with
the additional support grid enabled, the electronic convergence threshold
$10^{-6}$~eV, the non-spherical gradient correction inside the augmentation
spheres enabled, and the one-center occupancies mixed up to $l=4$, which the
$5d$ shell of tungsten requires.  Spin--orbit coupling is included
non-collinearly and dispersion is treated with D3(BJ).  The cell height is
30~\AA, giving about 20~\AA{} of vacuum above a slab of thickness
$\dslab$~\AA.  Under strain the interlayer distance was relaxed at every
point, and the scans run over $\pm2\%$, which brackets both crossovers with
calculated points.  Both strain series are referred to the nonpolar
$a_0 = \alatTwoH$~\AA, which puts the two slopes on one axis, comparable
with each other and with the nonpolar series of
Ref.~\cite{PaperI_valleyGAA}.  The polar registries have the larger $a_0$,
and their series therefore sits above the zero-field
landscape of Sec.~\ref{sec:A1} by the slope times that difference.  The code is VASP, version \vaspnew{} throughout except for the
symmetrization controls of Sec.~\ref{sec:fieldaxis}, which compare
\vaspold{} with \vaspnew{} and hold everything else fixed.

A one-variable sweep on these structures puts every quantity reported here
beyond the reach of the settings.  Taking the polar registry at zero field and changing one
parameter at a time, the valley offset moves by \convmesh~meV between
$12\times12$ and $24\times24$, by nothing between $24\times24$ and
$36\times36$, by \convencut~meV on raising the cutoff to 700~eV, by
\convediff~meV on tightening the convergence threshold to $10^{-8}$~eV, by
nothing on changing to tetrahedron smearing, and by \convcell~meV on
opening the cell to 35~\AA.  What remains is structural, and it is the force
criterion: two relaxations of the same structure under different cutoffs
land \convdww~\AA{} apart in the interlayer distance, and the
independently relaxed mirror registry lands \convmirror~\AA{} away, which
at the computed sensitivity of \dkgdww~meV\,\AA$^{-1}$ are
\convdwwmeV{} and \convmirrormeV~meV.

The hybrid pair was run on the same relaxed
geometry, cell, mesh, cutoff and dipole correction as the semilocal partner,
at $E_z = 0$ and $0.5$~V\,nm$^{-1}$ in both stackings.  The exchange
treatment is what changes between the legs.  The dispersion correction is
carried by the semilocal leg alone, which costs nothing here: D3(BJ) adds a
geometry-dependent energy and leaves the Kohn--Sham potential untouched, and
on one geometry the eigenvalues are unchanged by adding or dropping that
term.  A control on the nonpolar structure, one input run twice with the tag
and without, returns all \ivdweigN{} eigenvalues identical to the digits
printed.  Quantity by quantity: the interfacial potential goes from $\UFEdft$
to $\UFEHSE$~meV.  The field-induced rise in $\Delta_{K\Gamma}$ at
$0.5$~V\,nm$^{-1}$ goes from $\starkRPBE$ to $\starkRHSE$~meV in the polar
registry.  The unfitted dipole length goes from $\dKRpair$ to
$\dKHSEdirect$~nm, that is $\varepsilon_\perp$ from $\epsperpPBER$ to
$\epsperpHSE$.  In the nonpolar bilayer the $K$ doublet splits by
$\dEKtwoHHSEpair$~meV at HSE06 where PBE gives $\dEKtwoHPBEpair$~meV, the
field-induced rise goes from $\starkTwoHPBEnew$ to
$\starkTwoHHSEnew$~meV ($\starkTwoHpct\%$), and $\varepsilon_\perp$ goes
from $\epsperpPBEtwoH$ to $\epsperpHSEtwoH$.  Both legs of both pairs are
canonical structures, and the nonpolar pair shares the sampling of the
nonpolar field scan with the symmetrization disabled.  The two stackings
and the body therefore sit on one footing.

Equation~(\ref{eq:shiftfree}) is sensitive to how the layer weights are
defined.  They are charges projected onto atom-centered spheres of
radius $\rWpaw$~\AA{} on W and $\rSepaw$~\AA{} on Se, the default radii of
the projector-augmented-wave datasets used.
For each state the weight of a layer is the sum of the projected charge
over the one W and two Se of that layer.  Those sums are not unity --- they
come to $\projsumK$--$\projsumKhi$ for the states at $K$, the remainder
lying in the interstitial region --- so Eq.~(\ref{eq:tauznorm}) normalizes
by their total and not by their raw difference.

The two branches of a doublet are symmetrized before the inversion.  An
ideal two-level doublet satisfies
$\langle\tau_z\rangle_+ = -\langle\tau_z\rangle_-$ exactly; the computed
projections do not, violating that relation by up to \tauzviol\%, which is a measure of
the leakage that the normalization of Eq.~(\ref{eq:tauznorm}) cannot remove.
For the upper spin-orbit manifold of R-MX, for instance, the two branches give
$\langle\tau_z\rangle_{36} = \tauzupA$ and
$\langle\tau_z\rangle_{35} = \tauzloA$.  We symmetrize $|\langle\tau_z\rangle|$ within each same-spin doublet before inverting, which
uses both branches at once.

Four checks constrain the result.  First, $K$ is computed twice: spin--orbit
splits the valence states at $K$ into two manifolds \socKR~meV apart, and
because the layers of an R-stacked bilayer are parallel each manifold is
itself a same-spin layer doublet.  $U_{\rm FE}$ is electrostatic and $t_K$ spin independent, and the two must agree.  They do --- $\delta E_K = \dEKMXband$
and \dEKlo~meV, agreeing to \dEKspread~meV and giving $t_K^{\rm R}$ to within
\tKspread~meV.  With $n$ the band index, inverting
Eq.~(\ref{eq:zerofield}) on each manifold gives
\begin{equation*}
\begin{aligned}
n = 36,35:\;\; |\langle\tau_z\rangle_K| &= \tauzone,\\
  U_{\rm FE} &= \Uone~\text{meV},\quad t_K^{\rm R} = \tone~\text{meV},\\[2pt]
n = 34,33:\;\; |\langle\tau_z\rangle_K| &= \tauztwo,\\
  U_{\rm FE} &= \Utwo~\text{meV},\quad t_K^{\rm R} = \ttwo~\text{meV},
\end{aligned}
\end{equation*}
and hence, averaging over the two manifolds,
\begin{equation*}
U_{\rm FE} = \Uproj~\text{meV},
\qquad
t_K^{\rm R} = \tKproj~\text{meV},
\end{equation*}
of which the potential is the usable output and the hopping is not
(Appendix~\ref{app:tkroute}).  Second, the two registries are exact mirror
images: their total energies agree to the last computed digit and their layer
weights map into one another state by state.  Third, the model requires
$\delta E_K \geq U_{\rm FE}$ for any $t_K \geq 0$, which the extracted values
satisfy.

The fourth is independent of the band structure altogether.  The
planar-averaged potential gives a vacuum-level step across the slab of
$\vacstep$~meV for R-MX, exactly reversed for R-XM and unchanged between two
vacuum thicknesses.  That is an electrostatic quantity, obtained without
reference to any eigenvalue or projection. The $U_{\rm FE} = \UFEdft$~meV
extracted from the $K$ bands exceeds the vacuum step by \vacpct\% and the optically inferred value~\cite{Li2026_RWSe2} by \UFEpct\%.  The two determinations that use no band at $K$ both lie below the one that does.
The three span \threeway\%:
\begin{equation*}
\underbrace{\UFE}_{\text{optical}}
\;\;\; \underbrace{\vacstep}_{\text{vacuum}}
\;\;\; \underbrace{\UFEdft}_{\text{bands}}~\text{meV}.
\end{equation*}
The three quantities of Eq.~(\ref{eq:threeU}) need not coincide: the vacuum
step is the total dipole across the slab, the band inversion returns an
effective potential seen by the $K$ states, and the optical value is
inferred from a transition energy.  Identifying one $U_{\rm FE}$ among them
is a modeling choice.  That distinction appears in the magnitudes inside the slab as well,
where the band-derived step exceeds the planar-averaged potential difference
between the two W planes by a factor $\UoverPhi$: the $K$ state has W $d$
character and extends beyond the W plane over which the electrostatic one is
evaluated.

The electrostatic step has been measured on this system.  A graphene sensor
in a dual-gated stack reads the built-in interlayer potential of R-stacked
bilayer WSe$_2$ from the shift of the charge-neutrality point, giving
$\VPexpt(\VPexptsd)$~meV over several devices, and the calculation in that
work returns $\VPtheo$~meV~\cite{Wang2022_interfacialFE}.  That calculation
takes the step across the bilayer from the planar-averaged electrostatic
potential, at the semilocal functional and dispersion treatment of this work,
with the interlayer distance taken from bulk 2H and not relaxed.  Our
two readings of that profile, at the W planes and across the vacuum,
are $\dphiRzero$ and $\vacstep$~meV.  The measurement sits below both and
far below the band-derived $\UFEdft$~meV, which is the ordering
Eq.~(\ref{eq:threeU}) is about.

The $\Gamma$ point tests $t_\Gamma$ well and $U_{\rm FE}$ badly.
Appendix~\ref{app:bands} inverts the zone-center splitting for both
stackings and both polar registries and finds one $t_\Gamma$ to two per cent.  Thus, the \tGamma~meV of the equal-$d$ fit is a feature of that fit and not of the stacking.  The $\Gamma$ estimate of $U_{\rm FE}$ is \UFEgamma~meV, and that is ill
conditioning and not a valley-dependent electrostatic potential.  With
$t_\Gamma \gg U_{\rm FE}/2$ the layer polarization at $\Gamma$ is a small difference of two large projections.  Equation~(\ref{eq:zerofield}) is then ill conditioned for $U_{\rm FE}$ there and remains well conditioned for
$t_\Gamma$, which changes by under 1~meV across the whole plausible range of
$\langle\tau_z\rangle_\Gamma$.  Matching the $K$-point value would need
$|\langle\tau_z\rangle_\Gamma| = \tauzGneed$, while the calculation gives \tauzG,
a \tauzGshiftpct\% change in each raw weight, which the systematic
uncertainty of projections summing to only \projsumG{} allows.  We take
$U_{\rm FE}$ and $t_K$ from $K$ and $t_\Gamma$ from $\Gamma$, and keep the
$\Gamma$ estimate of $U_{\rm FE}$ as a caveat without averaging that value
in.

Normalizing the layer projections changes the answer.  Taking the bare
difference $W_{\rm top}-W_{\rm bot}$ in place of Eq.~(\ref{eq:tauznorm}) would return
$|\langle\tau_z\rangle_K| \simeq \tauzraw$, and hence a spurious
$t_K^{\rm R} \simeq \tKspurious$~meV.  That is squarely in the range that
would have suppressed the effect, and comes from a state that is in fact
almost completely layer-polarized:
\begin{equation}
\langle\tau_z\rangle_\nu = \frac{W_{\rm top}^{(\nu)}-W_{\rm bot}^{(\nu)}}
                                {W_{\rm top}^{(\nu)}+W_{\rm bot}^{(\nu)}} .
\label{eq:tauznorm}
\end{equation}

\section{Calculated band structures}
\label{app:bands}

Two load-bearing statements concern wavefunctions, not energies.  The first is that the $\Gamma$ edge is shared between the layers
and the $K$ edge lies on one of them.  The second is that the two polar registries are
mirror images with the layer labels exchanged.  Both can be read directly off the calculated bands.  The panels also give
$t_\Gamma$ with no fitting, and they show what the layer projections do
as the $K$ doublet is closed, which is the evidence behind
Fig.~\ref{fig:avoided}(c).

\begin{figure*}[tb]
  \includegraphics[width=0.95\textwidth]{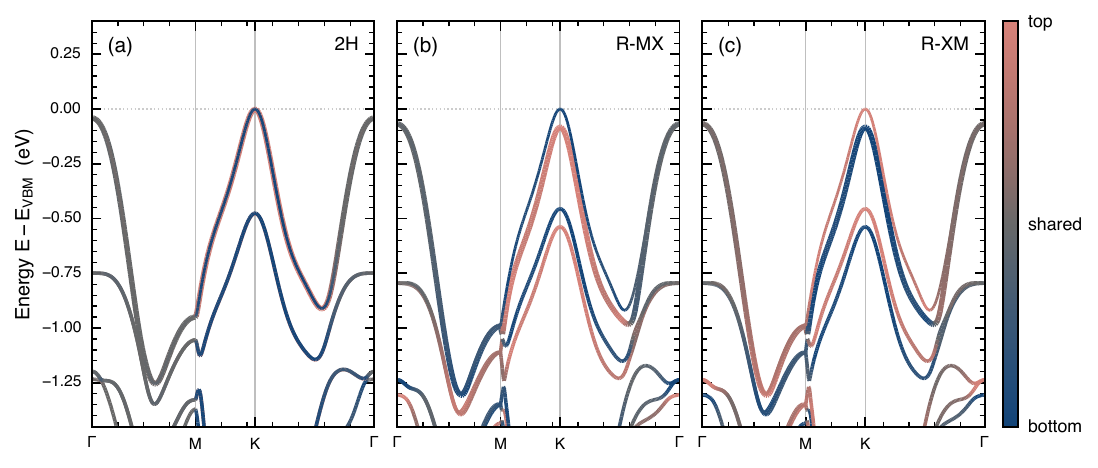}
  \caption{%
    Calculated valence bands at zero field, colored by the fraction of each
    state on the upper layer: blue is the lower layer, red the upper, and
    gray a state shared between them.  At $\Gamma$ the top valence states are
    shared; at $K$ they are polarized onto one layer, and (b) and (c) are
    mirror images with the layers exchanged.  In (a) symmetry makes the two
    $K$ branches exactly degenerate, and they lie on top of one another; the
    band step resolves them only to $\dEKtwoHband$~meV, which is the floor
    of that step and not a splitting.  The upper
    branch of the pair is drawn narrower in every panel so that the partner
    underneath stays visible.  The $K$-point splittings of the two polar
    registries are $\dEKMXband$ and $\dEKXMband$~meV for (b) and (c).%
  }
  \label{fig:bands}
\end{figure*}

Figure~\ref{fig:bands} gives the valence bands of the three stackings at zero
field along $\Gamma$--M--K--$\Gamma$, colored by the fraction of the state
on the upper layer: the nonpolar bilayer in (a), R-MX in (b) and R-XM in (c).  At $\Gamma$ the top valence states are shared almost
equally, $\wGshared$ and $\wGsharedb$ in the two registries and exactly one
half in the nonpolar bilayer, as a bonding or antibonding combination must
be.  At $K$ every branch is polarized onto one layer: in the two polar
    registries the edge has $\wKedge$ or more of the weight there, and the two registries place the edge on opposite layers.  The nonpolar bilayer is polarized at $K$ as well, because spin--layer
locking polarizes the branches without splitting them: symmetry leaves the
two degenerate and on opposite spins, and the $\dEKtwoHband$~meV by
which the band step separates them is numerical, while R-MX and R-XM give
$\dEKMXband$ and $\dEKXMband$~meV.  This is the hybridization contrast of
Sec.~\ref{sec:calibration} read off the wavefunctions and not inferred from
a Stark response, and the mirror relation of Eq.~(\ref{eq:mirror}) made
visible.

The zone center also determines $t_\Gamma$ with no fit.  The two Kramers pairs
there are the bonding and antibonding combinations of the same valence state,
split by $2\sqrt{t_\Gamma^2+(U^\Gamma/2)^2}$, and in the nonpolar bilayer the polar term is absent, leaving the splitting at $2t_\Gamma$ outright.  The
computed splitting is $\dEGtwoH$~meV, hence
\begin{equation*}
t_\Gamma^{\rm 2H} = \tGtwoHband~\text{meV}.
\end{equation*}
The polar side licenses the construction, since there the result can be checked against an independent determination.  Inverting that expression for R-MX, whose
splitting is $\dEG$~meV, returns $\tGRband$~meV, and R-XM gives
$\tGRbandXM$~meV.  The step entering there is $\Gamma$'s own, $\UGband$~meV, and not the $\UFEdft$~meV of $K$.  That choice changes the answer by only $\tGUsens$~meV, since $t_\Gamma$ is far larger than either.  The field scan
of Sec.~\ref{sec:A1}, which shares no input with the zone-center splitting,
returns $\tGR$~meV, the same value to the precision of either determination.  The consequence for the argument of
Sec.~\ref{sec:A1} is that $t_\Gamma$ is essentially the same in the two stackings.  Thus, the renormalization of the landscape by stacking is not a change in the interlayer hybridization but a change in $\bar E_K -
\bar E_\Gamma$ and in the relaxed geometry.

\begin{figure*}[tb]
  \includegraphics[width=0.95\textwidth]{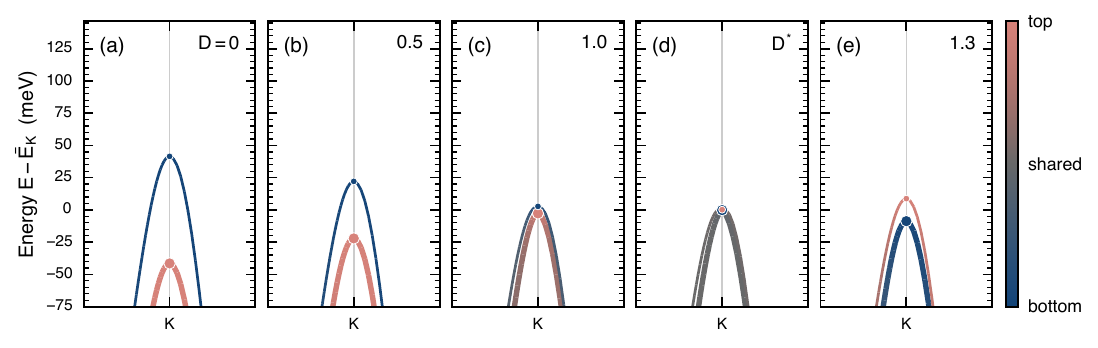}
  \caption{%
    The $K$ doublet of R-MX as the displacement field closes the doublet.  Each panel
    shows the same slice of the path, and energies are referred to the
    midpoint of the doublet at $K$, so that only the closing moves.  Panels
    are labeled by $D$ in V\,nm$^{-1}$, and the splittings at $K$ are
    $\dEKMXband$, $\dEKseriesa$, $\dEKseriesb$, $\dEKDstar$ and
    $\dEKpast$~meV.  Color is
    the layer weight as in Fig.~\ref{fig:bands}; circles are the states
    calculated at $K$ itself.  The branches stay polarized
    on opposite layers as the doublet closes, and in (e) they have exchanged
    layers relative to (a): the polarity reverses on crossing $D^{*}$.  In (d)
    the states away from $K$ are equal mixtures, because the on-site energies
    are degenerate there and any coupling then mixes them fully.  At $K$
    itself, where the coupling vanishes, they stay polarized.%
  }
  \label{fig:bandsfield}
\end{figure*}

Figure~\ref{fig:bandsfield} follows the $K$ doublet of R-MX out to and
through the compensation field, where the caption gives the five splittings.
The polarity of the doublet reverses on crossing that field, as
Eq.~(\ref{eq:eigen}) requires, and the reversal is the microscopic picture
of the displacement along the field axis.  The picture does not make $g_d$ antisymmetric between the
registries; that follows separately from Eq.~(\ref{eq:mirror}), and the two
statements concern the same landscape and different quantities.

Panel (d) is the closest approach to $t_K^{\rm R}$, because
$\delta E_K \geq 2t_K$ identically and the compensation field is where the
polar term is gone.  Taken at face value a splitting of $\dEKDstar$~meV would
cap the hopping at $\tKbandface$~meV, with no fitting and no extrapolation, while
Appendix~\ref{app:tkroute} obtains $\tKlimitcanon$~meV from the
self-consistent scan.  The splitting cannot be read that way, and we take the result instead as consistency with zero.  Run that
non-self-consistent band step
on the 2H zero-field doublet, which symmetry makes exactly degenerate, and the step returns $\dEKtwoHband$~meV; the $\dEKDstar$~meV found here lies below the resolution of the step that produced the value.  That floor is coarser than the $\dEKtwoHzero$~meV the self-consistent
scan gives for that 2H doublet in Sec.~\ref{sec:calibration}, because a
band calculation inherits whatever asymmetry the stored density has.  The
comparison here is with the band step, because the band step produced
$\dEKDstar$.

The layer weights in panel (d) show the same thing independently, and the
projection is well conditioned there.  At $D^{*}$ the two on-site energies
coincide, so any interlayer coupling mixes the branches completely and
$\langle\tau_z\rangle$ must vanish wherever $t$ is finite.  At just
$\mixdk$~\AA$^{-1}$ from $K$ the weights are already $\wmix$ and the
inversion returns
$t \simeq \tatmix$~meV, rising to $\tofkfar$~meV towards $M$ at
$0.077$~\AA$^{-1}$ and to $\tofkfarG$~meV towards $\Gamma$ at the same
distance.  The residual $\Delta_K/2$ stays between $\mixhalflo$ and
$\mixhalfhi$~meV across the window, which confirms that the polar
term really is canceled at this field.  At $K$ itself the branches keep their layer character, $\wKpol$ and $\wKedge$, and the inversion gives $t_K^{\rm R} = \tKDstar$~meV.

The way the coupling vanishes is the informative part.  The two directions out of $K$ are inequivalent, and we fit them separately over the calculated points within $0.05$~\AA$^{-1}$ of $K$, $\tfitptsKM$ towards M and $\tfitptsKG$ towards $\Gamma$:
\begin{equation}
t(\mathbf k) = v\,|\mathbf k - \mathbf K| + b ,
\label{eq:tnode}
\end{equation}
with $(v,b) = (\tslopeKM, \tinterKM)$ towards $M$ and
$(\tslopeKG, \tinterKG)$ towards $\Gamma$, in meV\,\AA{} and meV.  The slopes differ by a quarter, and a single fit over all $\tfitpts$ is the
worst of the three, but the two intercepts agree.  Each intercept is smaller, by at least a factor of \tinterfloor, than the
$\dEKtwoHband$~meV this band step returns for a doublet that symmetry makes
exactly degenerate, and the fits scatter by no more than
$\tnoderesid$~meV.  Thus, each direction places the coupling at $K$ at zero independently.  A coupling that happened to be numerically small would
be small in a neighborhood of $K$ and vary smoothly.  A linear node at one point of the zone is
the signature of a selection rule, and the rule is the three-fold one of
Sec.~\ref{sec:calibration}~\cite{Akashi2015_3RMoS2}.  The calculation here shows that the node is present and linear.  The
\polshiftdft~meV quoted in the text follows from
$t_K^{\rm R}=0$, which is both what the selection rule requires and what
the scan returns.  At the upper limit the conversion is still \effpct\% of
the ceiling.  The limit lowers the shift by less than the digits printed.

The bands plotted here are non-self-consistent runs along the path, on the
charge density of
the corresponding self-consistent calculation, with $\nbandk$ points and the
settings of Appendix~\ref{app:rparams}.  Across the polar series they sit
$\bandscf$~meV from the self-consistent values.  The nonpolar entries
differ by more, because those two calculations differ in mesh as well:
the band path gives $K$ splittings of $\dEKbandhalf$ and
$\dEKbandone$~meV at $0.5$ and $1.0$~V\,nm$^{-1}$ where the
self-consistent scan gives $\dekIsymhalf$ and $\dekIsymone$~meV.  The values quoted elsewhere are the self-consistent ones; the panels here are for reading,
not for measuring.  One field was recomputed here as a control on the
settings.  That repeat differs from the original in code version,
convergence threshold and band count, and the two agree to
$\bandrepro$~$\mu$eV on every one of the \nbandk{} $\times$ 11 eigenvalues.

\subsection{What the dipole-length inequality costs the electrical route}
\label{app:dgamma}

Section~\ref{sec:counting} reads the two dipole lengths and the two
interlayer steps off the doublets of one scan and gives the ratios.  The
polar series resolves both doublets over \nfieldsR{} fields, each obeying
$\delta E_\nu = 2\sqrt{t_\nu^2 + (\Delta_\nu/2)^2}$ with
$\Delta_\nu = sU^\nu + e d_\nu D$ to $0.02$~meV or better, and returns
\begin{equation}
  \frac{d_\Gamma}{d_K} = \dratioGK, \qquad
  \frac{U^\Gamma}{U^K}  = \UratioGK ,
  \label{eq:dratio}
\end{equation}
with $d_\Gamma = \dGband$~nm and $d_K = \dKbandfit$~nm, $U^\Gamma =
\UGband$~meV and $U^K = \UKband$~meV.  All four come from that one series.

The
electrical locator of Sec.~\ref{sec:electrical} needs the two valleys to be
canceled at the same field, and the two ratios in Eq.~(\ref{eq:dratio}) decide
whether they are.  They are close --- the $\Gamma$ valley is screened by
much the same factor whether the interlayer potential is polar or applied ---
but not equal.  The inequality displaces the zero of $g_d$ in
Sec.~\ref{sec:electrical}.  The $K$-valley zero is itself a second determination of the
compensation field, from the band structures alone: \nullK~V\,nm$^{-1}$,
consistent with the $\Dstardft$~V\,nm$^{-1}$ that the field scan gives through
$U_{\rm FE}/(ed_K)$, with no shared input.

\section{Ill conditioning of the layer-projection route to \texorpdfstring{$t_K^{\rm R}$}{tK(R)}}
\label{app:tkroute}

Equation~(\ref{eq:zerofield}) is exact, but is not equally well conditioned in the two outputs.  The interfacial potential $U_{\rm FE} = \delta E_\nu
\langle\tau_z\rangle_\nu$ is linear in the layer polarization and is robust.
The hopping is not: $t_\nu = \tfrac12 \delta E_\nu\sqrt{1-\langle\tau_z
\rangle_\nu^2}$ depends on $\langle\tau_z\rangle$ through a square root whose
derivative diverges as $\langle\tau_z\rangle \to 1$, which is precisely the
regime a nearly layer-decoupled valley occupies.  With the projection
convention of Appendix~\ref{app:rparams}, whatever leakage the
normalization there fails to remove is amplified by that divergence.  Applied to the $K$ doublet the inversion returns
$t_K^{\rm R} = \tKproj$~meV, and a leakage of under two per cent accounts for the whole of that value.

The band energies settle $d_K$ as well, and reject an unscreened dipole
outright.  The canonical scans measure $d_K = \dKR$~nm directly, an eighth
of the interlayer distance of that registry, \dWWR~\AA, and screening can
only shorten that length further.  Layer projections separate
the two more directly: at $D = 0.2$~V\,nm$^{-1}$ the top-layer weight of the
$K$ edge is \wtopzero{} for $t_K = 0$ but \wtopfifty{} for $t_K = 50$~meV.

\begin{figure*}[tb]
  \includegraphics[width=0.95\textwidth]{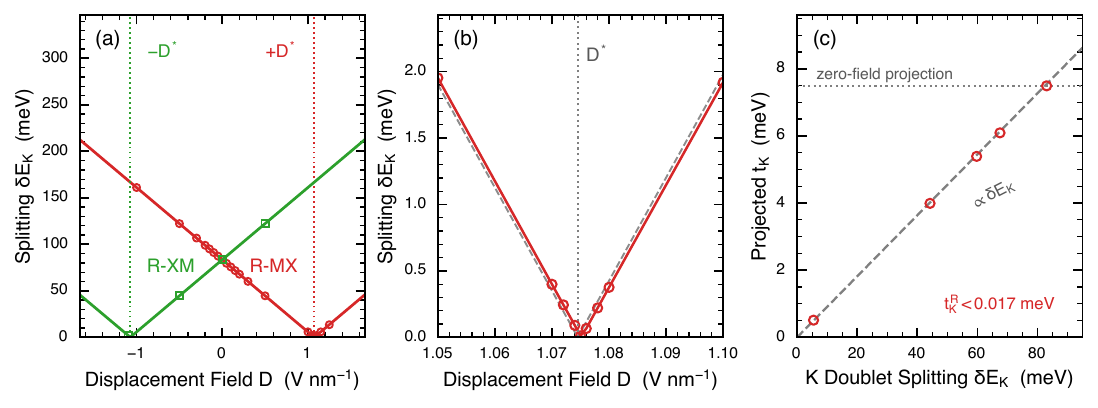}
  \caption{%
    Displacement-field dependence of the $K$-doublet splitting, and the resulting bound on $t_K^{\rm R}$.  Symbols are the calculated splittings of the
    field scan, curves are Eq.~(\ref{eq:eigen}) with the fitted parameters.
    No band is drawn for $t_K^{\rm R}$: the bound of
    Eq.~(\ref{eq:tKfield}) is a thirtieth of the width of the curves here.
    (a)~Both registries.  The polarization displaces the two curves in opposite directions along the field axis, and
    the splitting closes at $+D^*$ for R-MX and at $-D^*$ for R-XM, with
    $D^* = \Dstardft$~V\,nm$^{-1}$; at $D = 0$ the two coincide at
    $U_K^{\rm eff} = \UFEdft$~meV.  This is the mirror relation of
    Eq.~(\ref{eq:mirror}) written in a calculable quantity.
    (b)~The R-MX minimum, magnified.  The scan walks into the node, with
    \scannodepoints{} fields inside $0.005$~V\,nm$^{-1}$ of the compensation
    field, and closes the doublet to \dEKnode~meV there.  The dashed line is
    the cusp that $t_K^{\rm R} = 0$ would give.  No floor is drawn at
    $2t_K^{\rm R}$, because the bound on the hopping is below what the
    calculation resolves.
    (c)~Why the layer projection cannot be used instead.  Splitting
    dependence of the $t_K$ obtained by inverting the projection, one symbol
    per field, from the band structures of Appendix~\ref{app:bands}.  A hopping is a property of the Hamiltonian and
    would lie on a horizontal line; these lie on a ray through the origin
    (dashed, slope \projratio).  A fixed unassigned weight in the projection
    produces exactly that ray.  The dotted line is the zero-field value
    $\tKproj$~meV, the upper end of the same ray.  The solid line along the
    foot of the frame is the bound of Eq.~(\ref{eq:tKfield}), which on this
    axis is indistinguishable from zero.%
  }
  \label{fig:avoided}
\end{figure*}

The field scan bounds $t_K^{\rm R}$ without projections at all.  Under field the two
$K$ branches approach and separate again, and the separation at the closest
approach is $2t_K^{\rm R}$ regardless of where along the field axis that minimum lies.  That is an eigenvalue difference, and no layer assignment enters.  Figure~\ref{fig:avoided} shows the scan with the fitted curves.  We fit all \scanpointscanon{} fields, imposing $t_K \geq 0$ as the boundary it is, and the fit gives a central value at that boundary with an rms residual of \scanresidcanon~meV.  The
scan passes through the node, with
\scannodepoints{} fields inside $0.005$~V\,nm$^{-1}$ of the minimum, so the
depth is resolved and not extrapolated.

A central value at a boundary cannot carry a symmetric interval, so we
quote a one-sided limit from the profile likelihood, refitting
$U_{\rm FE}$ and $d_K$ at each fixed $t_K$:
\begin{equation}
t_K^{\rm R} < \tKlimitcanon~\text{meV} \quad (\tKlimitcl\%),
\label{eq:tKfield}
\end{equation}
the other registry giving the same limit independently.  The limit follows
from the scatter of the fit, \scanresidcanon~meV, which sits at the
\dEKtwoHzero~meV floor found independently on the nonpolar doublet that
symmetry makes degenerate.  The limit is therefore a statement about the
resolution of the calculation, and not a measurement of the hopping, which
the selection rule of Sec.~\ref{sec:twoforbid} sends to zero exactly.  The
bound is what the data give to a reader who declines to grant that rule,
and a better calculation would return a smaller one.  For the same reason
the text nowhere divides $t_\Gamma^{\rm R}$ by the bound: the quotient
$t_\Gamma^{\rm R}/t_K^{\rm R}$ has no finite value, and $>\tratiobound$ is
all that follows from the numerator and a bound on the denominator.  Against the room-temperature
thermal energy the bound is three orders of magnitude down, which is the
comparison that decides whether the hopping can do anything.  It excludes the
projection value outright.  The splitting can never fall below $2t_K^{\rm R}$, and $\tKproj$~meV imposes a floor of \twotKproj~meV.  At the
compensation field the polar term is gone, so a hopping of that size would
leave the doublet open by the whole \twotKproj~meV.  The scan closes that
doublet to \dEKnode~meV instead.  The layer projections at that field are moreover still fully polarized, which
a hopping of that size would not allow either.

The scan also diagnoses the projection directly.  Were the residual
$1-\langle\tau_z\rangle^2$ hybridization, the mixing would grow as the doublet closes,
and at the minimum the two branches would be equal mixtures of the layers.
Across the \projpoints{} R-MX band structures of Appendix~\ref{app:bands},
which span $\delta E_K$ from \projdEhi{} down to \projdElo~meV,
$|\langle\tau_z\rangle|$ does not change at all: the value is \projtauz{} at both ends.
The hopping inferred from the projection is proportional to the splitting,
$t_K \simeq \projratio\,\delta E_K$ to within \projratiopct\%.  A fixed
unassigned weight in the projection produces that proportionality, and a
matrix element of the Hamiltonian cannot
[Fig.~\ref{fig:avoided}(c)].  The zero-field value \tKproj~meV is the
upper end of that line, not a measurement.  We use Eq.~(\ref{eq:tKfield}) throughout and quote $\tKproj$~meV only as what the
projection route yields, for the benefit of anyone repeating that route.  The
hybridization contrast is reported as the bound in
Eq.~(\ref{eq:tKfield}) because a ratio whose denominator is consistent with
zero has no stable central value.

\section{Test of transferability from 2H to R}
\label{app:transfer}

Without the R-stacked structures, the only route to them is to assume that
the layer-symmetric parameters $\bar E_\nu$, $t_\Gamma$ and $d_\nu$ transfer from 2H, with the polar term $s\,U_{\rm FE}$ as the sole new ingredient
and $U_{\rm FE}$ taken from experiment~\cite{Li2026_RWSe2}.
The assumption is plausible, since the two registries differ by a lateral
shift with no change of interlayer separation and their total energies
differ by about \stackEdiff~meV per formula unit~\cite{He2014_stacking}.  We did not extend the assumption to $t_K$, for the symmetry reason given above. Under that assumption the R-stacked landscape is the 2H landscape rigidly displaced along the field
axis,
\begin{equation}
\Delta_{K\Gamma}^{(s)}(D) \simeq \Delta_{K\Gamma}^{\rm 2H}(D + s D^{*}),
\qquad D^{*} = \frac{U_{\rm FE}}{e\,d_K},
\label{eq:shift}
\end{equation}
rigidly, because the hypothesis puts the whole of the registry dependence
into the polar term and leaves every other parameter alone.  The direct
calculations of Sec.~\ref{sec:A1} test Eq.~(\ref{eq:shift}), and the way the relation fails is the informative part.

The two fields compared with experiment are of different kinds.  The
built-in field of \Dbuiltin~V\,nm$^{-1}$ quoted in
Ref.~\cite{Li2026_RWSe2} is $\varphi_0/d_0$, with $d_0 = \dzeroLi$~nm the
interlayer separation assumed there, and is an \emph{internal} field, whereas
$D^{*}$ is the external displacement field.  Converting with the
out-of-plane permittivity our own field scan implies,
$\varepsilon_\perp = \epsR$, gives \Dbuiltinext~V\,nm$^{-1}$ for the experimental value, compared with $D^{*} = \Dstardft$~V\,nm$^{-1}$.

What the comparison establishes is agreement on the \emph{dipole length}:
the screened $d_K$ reproduces the experimental internal field from the
optical $\varphi_0$.  The residual difference is the gap between the
computed and the optical interfacial potential, which
Eq.~(\ref{eq:threeU}) states and does not resolve.

Section~\ref{sec:A1} states the verdict: the registry mirror holds
exactly and the identification with a translated 2H landscape does not.
Here is the arithmetic.  Numerically, the transferred curve sits \rigidzero~meV from the directly
computed one at $D=0$ and departs from the computed curve by up to \rigidmax~meV over the
scanned range.  The departure is not spread evenly: the gap peaks near the compensation field, because the transferred model inherits the optical $\varphi_0$ and locates $D^{*}$ at \Dstartrans~V\,nm$^{-1}$.  The computed value is \Dstardft.  A rigid translate along the field axis removes part of that, leaving
\rigidshift~meV, and allowing a vertical offset as well brings the residual to \rigidboth~meV.  That last value is the informative one.  \emph{The shape
transfers; the two scales do not.}  The shapes themselves offer no second
discrepancy to quote: the selection rule of Sec.~\ref{sec:calibration} forces $t_K^{\rm R}(K)$ to zero, and both minima are cusps with no curvature ratio between them.

Equation~(\ref{eq:newstory}) states the two parts.  Comparing with the
transferred model quantifies the assumption of Sec.~\ref{sec:landscape},
provided the two are compared like for like.  The
polar-term $K$ shift of Eq.~(\ref{eq:shiftvstk}) is \polshiftdft~meV, while
a decoupled valley with the optical $U_{\rm FE}$ would give \UFEhalf~meV.  The assumption changes the answer by \shiftpct\% there.  The
\emph{landscape} difference is closer: the transferred model gives \polshift~meV for $\Delta_{K\Gamma}^{\rm R}(0)-\Delta_{K\Gamma}^{\rm 2H}(0)$,
and the two computed structures \dkgdiffdft~meV, a gap of
\dkgdiffresid~meV.  That closeness is a cancellation and not a success.
Taking the optical potential costs \polshiftgap~meV in the polar term, and
the same model omits the stacking renormalization, which lowers the $K$ edge
by \EbarKrenorm~meV.  The two act against one another and
\dkgdiffresid~meV survives.  The $\Gamma$ edge takes no part: the two
models agree there to \GextraR~meV, which is the same statement as the two
stackings sharing their $\Gamma$ hybridization to \tGbandpct\%.  Thus, the assumption had the sign, the symmetry and the order of magnitude right, and had the two
underlying scales wrong in opposite directions, which is a more useful
statement than either ``confirmed'' or ``refuted''.  The dipole length is no part of that hypothesis: $d_K = \dKR$~nm comes from
a field scan of both polar registries, which Sec.~\ref{sec:fieldaxis}
shows the nonpolar bilayer gives to \dagreepct\%, and the compensation
field $D^{*} = \Dstardft$~V\,nm$^{-1}$ follows from that length.

\section{Gate architectures as regions of one plane}
\label{app:gates}

The state of the channel is a function of $(p,D)$ alone, whatever produced that state.  The three gate architectures differ only in which part of that plane each can reach.  Table~\ref{tab:gates} compares them.

Two rows need a value the table has no room for.  A single gate reaches
$D^{*}$ only at $\psinglestar \times 10^{13}$~cm$^{-2}$, where the
differential response is weakest, and the landscape a wrapped gate operates
on still differs by \dkgdiffdft~meV between R and 2H.

\begin{table*}[tb]
  \caption{The three gate architectures as regions of one plane.  A wrapped
  gate reaches the line $D=0$, a single gate the line
  $D=qp/2\varepsilon_0$, and only a dual gate the interior, so only a dual
  gate varies the field at fixed density.  Results obtained here specialize
  to the other two architectures by restriction to their lines.}
  \label{tab:gates}
  \begin{ruledtabular}
  \begin{tabular}{llll}
   & Gate-all-around & Single gate & Dual gate \\
  \colrule
  Electrodes            & one, wrapped
                        & one, one face
                        & two, independent \\
  Field at the channel  & $D=0$ structurally
                        & $D = qp/2\varepsilon_0$, pinned to $p$
                        & $D = \varepsilon_{ox}(V_t-V_b)/2t_{ox}$ \\
  Region of $(p,D)$    & the line $D=0$
                        & the line $D=qp/2\varepsilon_0$
                        & the plane \\
  Largest $D$ reached   & none
                        & \Dsinglepct\% of $D^{*}$ at $10^{13}$~cm$^{-2}$
                        & any, at fixed $p$ \\
  Polarization enters   & inside a fixed landscape
                        & along one line of that plane
                        & as a displacement that can be undone \\
  Observable            & occupation only
                        & the two responses, superposed
                        & $\chi_v^{(n)}$ and $\chi_v^{(D)}$ separately \\
  \end{tabular}
  \end{ruledtabular}
\end{table*}

\section{Robustness of the predictions}
\label{app:robust}

Every quantity that enters the predictions is determined by calculation in
Secs.~\ref{sec:calibration}--\ref{sec:strain}.  Some of the inputs remain uncertain, and the predictions divide into those
that survive them and those that do not.

Through Eq.~(\ref{eq:shiftvstk}) the magnitude of the whole effect would
rest on $t_K^{\rm R}$, which is why the scans below keep it as a variable.
Of the inputs scanned here, only the R-stacked field scan of
Sec.~\ref{sec:A1} bounds that hopping without invoking the three-fold
selection rule, and the spin--layer-locking argument that forces $t_K=0$ in
2H does not transfer to R stacking.  That bound,
Eq.~(\ref{eq:tKfield}), sits three orders of magnitude under the thermal
energy, so the dependence survives here as a stress test and not as an
uncertainty.

Two of the scanned entries come from one experiment and are mutually
inconsistent, and the source can be located.  Neither $\varphi_0$ nor $\Delta_{\rm v}$ is measured as an energy:
Ref.~\cite{Li2026_RWSe2} obtains both by multiplying the same assumed
separation $d_0 = \dzeroLi$~nm by a field, and the two fields differ,
\Dbuiltin~V\,nm$^{-1}$ for $\varphi_0$ and \DvField~V\,nm$^{-1}$ for
$\Delta_{\rm v}$.  Their ratio is precisely the \liinconsist\% of
Sec.~\ref{sec:liscales}.  The inconsistency lies between two determinations of the internal field inside one experiment.  No choice of
$t_K$ removes that gap, and none of this is evidence that the two-level
relation fails.  One possibility the published values cannot settle is an
excitonic binding correction in the optical transition behind $\varphi_0$
that the single-particle splitting $\Delta_{\rm v}$ does not include.

We use the degeneracy-weighted ratio $r_N = \rN$ of
Ref.~\cite{PaperI_valleyGAA} and not the bare mass ratio
$m^*_\Gamma/m^*_K = \rmass$ of Ref.~\cite{PaperII_transconductance}.
Susceptibilities quoted in one of the two are not comparable
with those quoted in the other.  In the gate-all-around geometry of
Ref.~\cite{PaperII_transconductance} at 300~K, on the landscape computed
here, the two conventions locate the peak at $\chivpeak$ and
$\chivpeaklegacy$~V$^{-1}$.  The prefactors differ
by exactly $g_K/g_\Gamma = 2$; the peaks do not, because $\chi_v$ is not
linear in $r_N$.  That reference runs on an offset of $\dkgpaperII$~meV
where the canonical structure gives $\dkgTwoHPBE$~meV.  On the smaller
offset, in the bare-ratio convention, the calculation here returns
$\chivpaperII$~V$^{-1}$ against the $\chivpaperIIpub$~V$^{-1}$ printed
there.  The two treatments
agree once the landscape matches.  Two reasons settle the choice.  First, Ref.~\cite{PaperI_valleyGAA}
defines $g_K = 4$ and $g_\Gamma = 2$.  Second, the
layer-resolved treatment makes the counting explicit: the $\mathcal{PT}$ doublet at $K$ \emph{is} the layer doublet, keeping both
branches leaves a degeneracy of two for each, and summing them returns
$r_N$.

Two numerical decompositions belong here, where
Sec.~\ref{sec:whygd} quotes only their outcome.  The first is why
2H and R are so alike in direct current.  Going from 2H to R raises
$\Delta_{K\Gamma}$, which drives holes out of the heavy $\Gamma$ valley and
would by itself cut $f_\Gamma$ by \fGlanddown\%; the change also splits the $K$ layer doublet, by $\delta E_K = \dEK$~meV where 2H has the doublet degenerate, which
lifts half of the $K$-edge density of states out of the thermal window and
raises $f_\Gamma$ by \fGsplitup\%.  The net change is $\dfGpct$\%.  Less of
even that reaches the current: the residual $\Gamma$ depletion changes
the effective mobility by \dmuPct\% and the carrier density at fixed gate
bias by \dpPct\%, and the two partly cancel in the product.

The second is the corollary on $t_K^{\rm R}$.  A large $t_K^{\rm R}$ weakens
the polarization-induced shift, and also pushes the lower $K$ subband out
of the thermal window: the share of the $K$ population in that subband falls from
\KlowshareZero\% at $t_K^{\rm R}=0$ to \KlowshareFifty\% at 50~meV.  The
effective $K$-valley density of states drops with the share and holes move into $\Gamma$.  Both effects raise $f_\Gamma$.  The mobility falls by \dmuPctmax\%, and the difference in current between 2H and R \emph{widens},
to \dIDpctmax\%.  A large measured direct-current contrast would
indicate a weak landscape effect, and a clean $g_d$ signal with little
contrast a strong one, which makes the pair an experimental check on the
vanishing of $t_K^{\rm R}$.

Several practical points take a quantitative answer, because they bear on
how the measurement degrades and not on whether the measurement exists.  The
four that follow displace the zero of $g_d$ that Sec.~\ref{sec:electrical}
locates, or remove the zero altogether, and the two after them leave the zero
where it is.

\emph{Gate-stack asymmetry.}  The gate stack has to be symmetric.  When the
two capacitances differ, $V_d$ drives the common-mode electrostatics at first
order, through $(C_t-C_b)V_d/(C_t+C_b)$, and the parity is broken at that
order.  The current then rises monotonically through $-sD^{*}$ and there is
no zero to find.  The parity in $D$ itself holds on any stack, and an asymmetric device recovers the zero only if the common mode is compensated
as $V_d$ is swept.

\emph{Layer-dependent transport.}  The mirror about $-sD^{*}$ leaves every
band energy alone but exchanges the two layers, $w_{\rm top} \to 1-w_{\rm
top}$.  The carrier imbalance is then odd in $D+sD^{*}$, and the current is
even only where transport cannot distinguish the layers.  That is the
field-symmetric-scattering idealization of Sec.~\ref{sec:transport}, and the
idealization matters more here than at $D=0$, where the mirror maps each
layer onto itself.  A fractional difference $r$ between the two layers'
mobilities displaces the zero by $\mulayerslope\,r$ of $D^{*}$.  A
\mulayer\% asymmetry, such as two gate dielectrics of unequal quality would
produce, displaces the zero by \mulayershift\%, the largest of the
displacements collected here.

\emph{A built-in interlayer step.}  A device leaves a layer-asymmetric
offset at $D = 0$ from several sources: different dielectrics above and
below, gate metals with different work functions, fixed charge in one of
the oxides, and a substrate that dopes or binds the adjacent layer.  A
complete treatment would give that offset a valley index, since a layer
potential of environmental origin is felt by the two valleys in their own
proportions, as an applied $D$ is through $d_\nu$; writing one value for
both costs nothing here, the term being zero in every calculation
reported.  In Eq.~(\ref{eq:Delta}) $U_{\rm env}$ enters without the
registry sign $s$ and moves the zeros of the two registries one way.  The
mirror breaks, whereas the valley mismatch of Sec.~\ref{sec:electrical}
and the mobility asymmetry above move the zeros oppositely and leave the
mirror intact.  The displacement is \uenvslope\% of $D^{*}$ per meV,
putting \uenvmev~meV of built-in asymmetry at \uenvshift\%, between the
two effects above.

\emph{Spatial inhomogeneity.}  Treating the channel as patches conducting in
parallel, each with a local interfacial potential, the zero of the ensemble
sits at the field that cancels the \emph{mean} polarization.  A symmetric
spread leaves the zero at $-sD^{*}$ to the precision of the calculation at
every width up to $\pm\gdnullspread\%$, and a skewed distribution displaces
the zero to the weighted mean and no further, to within \gdnullmeanresid\%
over the distributions we have tried.  Thus, disorder in the local compensation field broadens the transition without displacing the center, and
where disorder does displace the center, the displacement is the device
average.  A macroscopic measurement should report that average in any case,
and the statement is the same as the domain average of
Sec.~\ref{sec:discussion} for a continuous distribution.

\emph{Contact resistance.}  A series contact resistance $R_c$, summed over
both contacts, does not displace the zero of $g_d$ at all: the measured current is a monotonic function of the channel conductance,
which is even about $-sD^{*}$, and any function of an even function is even.  The derivative vanishes at $-sD^{*}$ to
machine precision for every $R_c$ tried.  The amplitude falls as the square
of the divider ratio, since both the current and the derivative pick up one
factor of that ratio.  Against the channel resistance at the operating
point, $R_{\rm ch} = \Rchannel$~k$\Omega$, the peak response is
$\RcHalf$ of the ideal at $R_c = R_{\rm ch}/2$, $\RcOne$ at $R_c = R_{\rm
ch}$ and $\RcFive$ at $R_c = 5R_{\rm ch}$, within a few percent of
$(1+R_c/R_{\rm ch})^{-2}$ throughout.  The figure of merit is the
ratio to the channel resistance and not $R_c$ itself.

\emph{Strong accumulation.}  Section~\ref{sec:readpoint} argues for
operating near subthreshold.  The fractional response $|g_d|/I_D$ falls
from \gdfracLo{} to \gdfracHi~V$^{-1}$ as the density goes from $\pLow$ to
twice $\pHi$.  Biasing hard does more than weaken the signal: the direct-current contrast falls and reverses,
crossing zero by $\contrastFlip\times10^{13}$~cm$^{-2}$.  The crossing appears only once
the two valleys have appreciably different mobilities, and at the 300~K read
point of Sec.~\ref{sec:readpoint} the crossing traverses the whole range scanned above once the ratio exceeds about $\contrastFlipRatio$.  Below that
ratio the contrast is already positive at low density and there is nothing
to cross.  In the equal-mobility limit the contrast stays positive and
decays towards zero.  The threshold depends on the operating point: the
crossing density falls as the temperature rises, and by a comparable amount
as the low-density end is pushed deeper into subthreshold.  The differential
bias cancels from a ratio of currents and does not enter.  Both estimates
available here sit above the threshold, the computed masses giving $\rdosR$
and the photoemission $K$ mass $\massratioArpes$.  Neither is a transport
measurement: both convert masses into a mobility ratio by assuming the two
valleys share a scattering time, the modeling choice stated in
Sec.~\ref{sec:transport}.  The zero of $g_d$ is unaffected at every density
and every ratio examined.

The masses can be checked from outside, and on this system directly.  Angle-resolved photoemission on 3R bilayer WSe$_2$ gives a hole mass
at $K$ of $\mKarpes \pm \mKarpeserr\,m_0$~\cite{Chapuis2026_3RW110}, larger
by \mKarpespct\% than the $\mKR\,m_0$ computed here.  A second photoemission study, on chemical-vapour-deposited bilayers, gives $\mKcvd\,m_0$ for the lighter of the two spin-split bands at $K$~\cite{Mahmoudi2023_3Rbilayer}. Magnetotransport agrees: Shubnikov--de Haas oscillations in a gated 2H
bilayer give $\mKsdh\,m_0$~\cite{Fallahazad2016_mass}, above the
$\mstarK\,m_0$ computed for that stacking.  Two techniques with unrelated systematics sit together and above the calculation, which leaves
photoemission resolution unable to account for the difference alone.
We have not tried to account for the difference.  Substrate strain and many-body
renormalization both remain candidates.  The consequences are small: adopting the measured value
changes $r_N$ and $f_\Gamma$ by about a sixth, $|g_d|$ by \gdmasspct\% and the
read-out contrast by less, and leaves $D^{*}$ and the parity of
Eq.~(\ref{eq:gdnull}) where they were, because neither depends on a mass.

The functional divides the predictions in two.  The matched pair of
Appendix~\ref{app:rparams} leaves the interfacial potential and the field
response almost untouched, at $\UFEHSEpct\%$ and $\starkRpct\%$, changes the dipole length by $\dratioHSE\%$, from $\dKR$ to $\dKHSEdirect$~nm, and the
compensation field from $\Dstardft$ to $\DstarHSEnew$~V\,nm$^{-1}$, and
the landscape difference from $\dkgdiffPBEpair$ to $\dkgdiffHSE$~meV.  In contrast, the absolute offset changes by $\dkgshiftR$~meV, and with the offset the room-temperature occupation by a factor of $\fGratio$,
from $f_\Gamma \simeq \fGRPBE$ to \fGRHSE.  Predictions that rest on the
occupation inherit the full PBE--HSE06 band.  Those that rest on
the polarization-induced shift, on the mirror relation of
Eq.~(\ref{eq:mirror}) or on the nonpolar zero do not.

The mobility ratio is the one input here that is not a first-principles
quantity at all, and is the strongest influence on the \emph{magnitude} of
$g_d$.  Sweeping the ratio from $1$ to $10$, together with $t_K^{\rm R}$ over
$0$--$50$~meV, changes $g_d$ at $D = 0$ and $V_c = 0.30$~V over the whole range
$\gdbandlo$ to $\gdbandhi$~$\mu$A\,V$^{-1}$, and at the equal-mobility end of that sweep the scan reverses the sign of $g_d$ itself.  The hopping range here is a stress test and not a statement of
uncertainty: the bound of Eq.~(\ref{eq:tKfield}) is $\tKlimitcanon$~meV.
However, the range cannot touch the sign \emph{difference} between the registries.
At $D = 0$ the two are exact negatives of one another at every point of that
scan, by Eq.~(\ref{eq:mirror}).  That is why the read-out claim is framed as a
reversal between registries and not as a sign of $g_d$, and that is the one durable statement this scan produced.

\section{Systematics and limits of the thermal route}
\label{app:arrsys}

The curvature of an Arrhenius plot of the valley occupation comes from the
$K$ doublet alone.  Section~\ref{sec:arrhenius} states the result, and this
appendix gives the derivation, the systematics and the limits.

\begin{figure*}[tb]
  \includegraphics[width=0.95\textwidth]{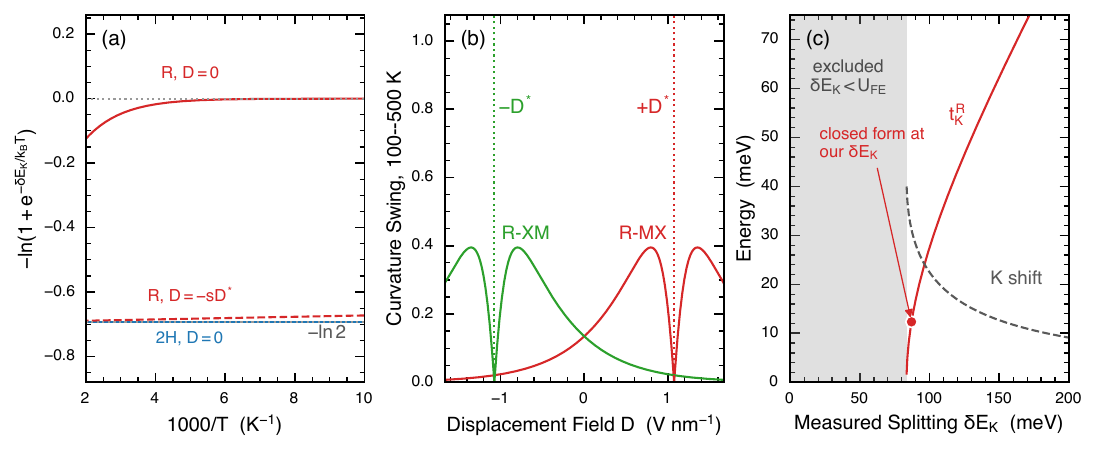}
  \caption{%
    Within the temperature-independent-band model, the $K$ layer doublet is
    the only source of curvature in a valley Arrhenius plot, which makes
    $\delta E_K$ measurable.  The systematics in the text treat the ways that
    model relaxes.
    (a)~The curvature term of Eq.~(\ref{eq:arrhenius}) as a function of the
    Arrhenius abscissa, for 2H at zero field and for R at zero field ($\dEK$~meV) and at
    the compensation field ($\dEKDstar$~meV).
    For the 2H bilayer at zero field, where $\delta E_K$ vanishes by symmetry,
    the curvature is pinned at $-\ln 2$ at every temperature.  The two R
    curves interpolate towards zero as the lower branch freezes out.
    (b)~Displacement-field dependence of the swing of that term over
    $\Tarrlo$--$\Tarrhi$~K.  The curvature is minimized at the compensation field $D = -sD^{*}$, where the
    applied field cancels the built-in polarization, and the minimum sits at
    opposite fields for the two registries.  \emph{The position is the
    observable, not the depth.}  The minimum closes completely, because the
    selection rule puts $t_K^{\rm R}(K)$ at zero and the scan finds the
    hopping there.  The bound of Eq.~(\ref{eq:tKfield}) would open the minimum to
    $\arrswingupper$, and the splitting the direct calculation gives leaves
    $\arrswingdirect$; both are below any practical resolution.  The flanking maxima occur
    where $\delta E_K \sim k_BT$.
    (c)~Inversion of a measured splitting, Eq.~(\ref{eq:tKfromdE}), for the
    interlayer hopping and for the polarization-induced $K$ shift.  The shaded
    region $\delta E_K < U_{\rm FE}$ is inaccessible for any $t_K \geq 0$ and
    would indicate that the two experiments disagree.  The field-scan bound,
    $t_K^{\rm R} < \tKlimitcanon$~meV, is not drawn: on this axis it is the
    axis.%
  }
  \label{fig:arrhenius}
\end{figure*}

In the non-degenerate limit the four subbands of Sec.~\ref{sec:doublet} give,
with energies measured from the upper $K$ edge,
\begin{equation}
\ln\frac{f_\Gamma}{1-f_\Gamma}
  = \ln 2r_N - \frac{\Delta_{K\Gamma}}{k_BT}
  - \ln\!\left(1+e^{-\delta E_K/k_BT}\right).
\label{eq:arrhenius}
\end{equation}
The factor of two counts the two $K$ branches, which the last term then
reweights.  When the doublet is closed, $\ln 2$ and the last term
cancel and Eq.~(\ref{eq:arrhenius}) reduces to the single-edge form that defines $r_N$
in Eq.~(\ref{eq:rN}).
The first two terms are a constant and a term linear in $1/T$, which leaves
\emph{the $K$ doublet as the only source of curvature} in this
temperature-independent-band model.  The last term varies from $-\ln 2$ when $k_BT \gg \delta E_K$ to
zero once the lower branch freezes out, a swing of \arrswingzero{} over
$\Tarrlo$--$\Tarrhi$~K for a decoupled valley
[Fig.~\ref{fig:arrhenius}(a)].

The minimum locates $D^{*}$ directly, because the position is independent of
the parameters.  Since $\Delta_K^{(s)}(D) = sU_{\rm FE} + ed_KD$, the layer
term vanishes at $D = -sD^{*}$ whatever the parameters, so the curvature is
minimized there [Fig.~\ref{fig:arrhenius}(b)].  Finding the minimum in data
therefore calls for an interpolation and no model.  The \emph{position} gives
the ratio $D^{*} = U_{\rm FE}/ed_K$ and nothing more.  Combined with a
separately measured $U_{\rm FE}$, the position yields the dipole length that
Sec.~\ref{sec:fieldaxis} obtains from two independent calculations.  The
\emph{depth} would in principle give $t_K^{\rm R}$, because the splitting
bottoms at $2t_K^{\rm R}$ and not at zero.  In practice the depth does not,
because $t_K^{\rm R}$ is too small.  At the splitting the direct calculation
gives, the residual swing is only \arrswingdirect{}, and the bound of
Eq.~(\ref{eq:tKfield}) permits no more than \arrswingupper{}.  The direct
calculation, the fit and the selection rule agree that the plot is
straight.  The
practical signature is therefore the opposite of a measurable depth: at
$D = -sD^{*}$ the Arrhenius plot is straight, and straight to within the
resolution of the method.  Neither the position nor the depth uses a layer projection.  That matters
because $\sqrt{1-\langle\tau_z\rangle^2}$ is steep near
$|\langle\tau_z\rangle| = 1$, and that steepness made the projection route
unusable in Sec.~\ref{sec:whichtK}.

The curvature cannot be imitated by the obvious systematic.  $\Delta_{K\Gamma}$ is itself
temperature dependent through thermal expansion and phonon
renormalization.  Writing $\Delta_{K\Gamma}(T) = \Delta_0 + aT$ turns the second term of
Eq.~(\ref{eq:arrhenius}) into $-\Delta_0/k_BT - a/k_B$, and the constant piece
changes only the intercept.  Only a \emph{nonlinear} thermal shift enters the
curvature.

Given $\delta E_K$, the splitting inverts for the hopping,
\begin{equation}
t_K^{\rm R} = \tfrac{1}{2}\sqrt{\delta E_K^2 - \varphi_0^2}, \label{eq:tKfromdE} \end{equation} in which $\varphi_0$ is the optically inferred interfacial potential of Ref.~\cite{Li2026_RWSe2}. Equation~(\ref{eq:tKfromdE}) comes with a consistency check: $\delta E_K \geq \varphi_0$ for any $t_K \geq 0$, and a measured splitting below $\varphi_0$ would mean the two experiments disagree,
not that the hopping is negative.  Figure~\ref{fig:arrhenius}(c) shows the inversion and the region excluded.

Four limitations bound the method.  The curvature vanishes in both limits
of Eq.~(\ref{eq:arrhenius}), and the method loses sensitivity above
$\delta E_K \approx \arrusable$~meV.  The two flat cases differ by an
intercept of $\ln 2$ whatever $r_N$ is, so that difference alone is usable.
Reading either intercept in absolute terms instead requires $r_N$, and hence
the effective masses, to be known independently.  The derivation
assumes non-degenerate statistics, and so shares the low-density operating
point identified in Sec.~\ref{sec:readpoint} and fails in strong
accumulation for the same reason.  Finally, $f_\Gamma$ is not measured
directly: extracting $f_\Gamma$ from the effective mobility, as in
Ref.~\cite{PaperII_transconductance}, introduces the mobility ratio and the
temperature dependence of that ratio.  A capacitance measurement is cleaner, because the lower $K$ branch adds a
step to the density of states.  The position of the zero comes from the field sweep and not from the
extraction.


\bibliographystyle{apsrev4-2}
\bibliography{references}

\end{document}